\documentclass[fleqn,10pt]{wlscirep}
\usepackage[utf8]{inputenc}
\usepackage[T1]{fontenc}
\usepackage{verbatim}
\title{Two-parameter Hong-Ou-Mandel dip}

\author[1,2]{Yu Yang}
\author[1,*]{Luping Xu}
\author[3]{Vittorio Giovannetti}
\affil[1]{School of Aerospace Science and Technology, Xidian University, Xi'an 710126, China}
\affil[2]{Scuola Normale Superiore, I-56126 Pisa, Italy}
\affil[3]{NEST, Scuola Normale Superiore and Istituto Nanoscienze-CNR, I-56127 Pisa, Italy}
\affil[*]{xidian\_lpx@163.com}

\begin{abstract}
A modification of the standard Hong-Ou-Mandel interferometer  is proposed which allows one to replicate the celebrated coincidence dip in the case of two-independent  delay parameters.
In the ideal case where such delays are sufficiently stable with respect to the mean wavelength of the pump source, properly symmetrized  input bi-photon states allow one to pinpoint their values  through the identification of a zero in the coincidence counts, a feature that cannot be simulated by semiclassical inputs having the same spectral properties. Besides, in the presence of fluctuating parameters the zero in the coincidences is washed away: still the bi-photon state permits to recover the values of parameters with a visibility which is higher than the one allowed by  semiclassical sources. The detrimental role of loss and dispersion  is also analyzed and an application in the context of  quantum positioning is presented.  
\end{abstract}
\begin{document}

\flushbottom
\maketitle
\thispagestyle{empty}

\section*{Introduction}
In a conventional Hong-Ou-Mandel (HOM) interferometer~\cite{HOM1,PhysRevLett.59.2044,HOM3,MREV}   photo-counting coincidences  are detected at the output ports of a 50:50 beam splitter that coherently mixes two multi-frequency optical waves coming from two separate spatial modes. The sensitivity of the registered signal is directly related with the indistinguishability of the interacting light fields ~\cite{IND1,IND2,SPDC2,IND3,IND4,IND5} making it a useful tool in a variety of information and measurement processing. In particular, using as input two properly correlated multi-frequency photons~\cite{SPDC,SPDC1,SPDC2}, the coincidences counts $R$ exhibits a unique minimum   $R=0$ (Mandel dip) when the difference  $\Delta \ell$ between the path length followed by the two  waves is zero, with a  width that is representative of the coherence time of the bi-photon state. Exploiting this feature, the HOM interferometer can  be turned into a  sensor for phase drifts or displacements measurements~\cite{SENSOR,SENSOR1} with possible applications in clock-synchronization ~\cite{CLOCK1,CLOCK2} and coordinates recovering~\cite{Bahder:2004} procedures.
Our main goal is verifying  the possibility of constructing a multi-parameter version of original HOM effect, i.e. a generalization of the HOM architecture where the absence of two-photon coincidences at the output of the interferometer is in a one-to-one correspondence with the case where more than one spatial coordinate have been set equal to zero contemporarily. Considering that  the coincidence counts $R$ at the output of the HOM setup is expressed as a one-dimensional convolution integral of the two-dimensional frequency spectrum of the bi-photon (BP) state, it is natural to expect that this task could be realized at least for two independent parameters, even though no solutions have been proposed so far.  Having clarified this aspect our secondary goal is to find applications for this effect taking inspiration from those proposed for the original HOM scheme, namely
as a tool for discriminating quantum from classical sources and as a tool for sensing.

\begin{figure*}
	\includegraphics[width=0.5\columnwidth]{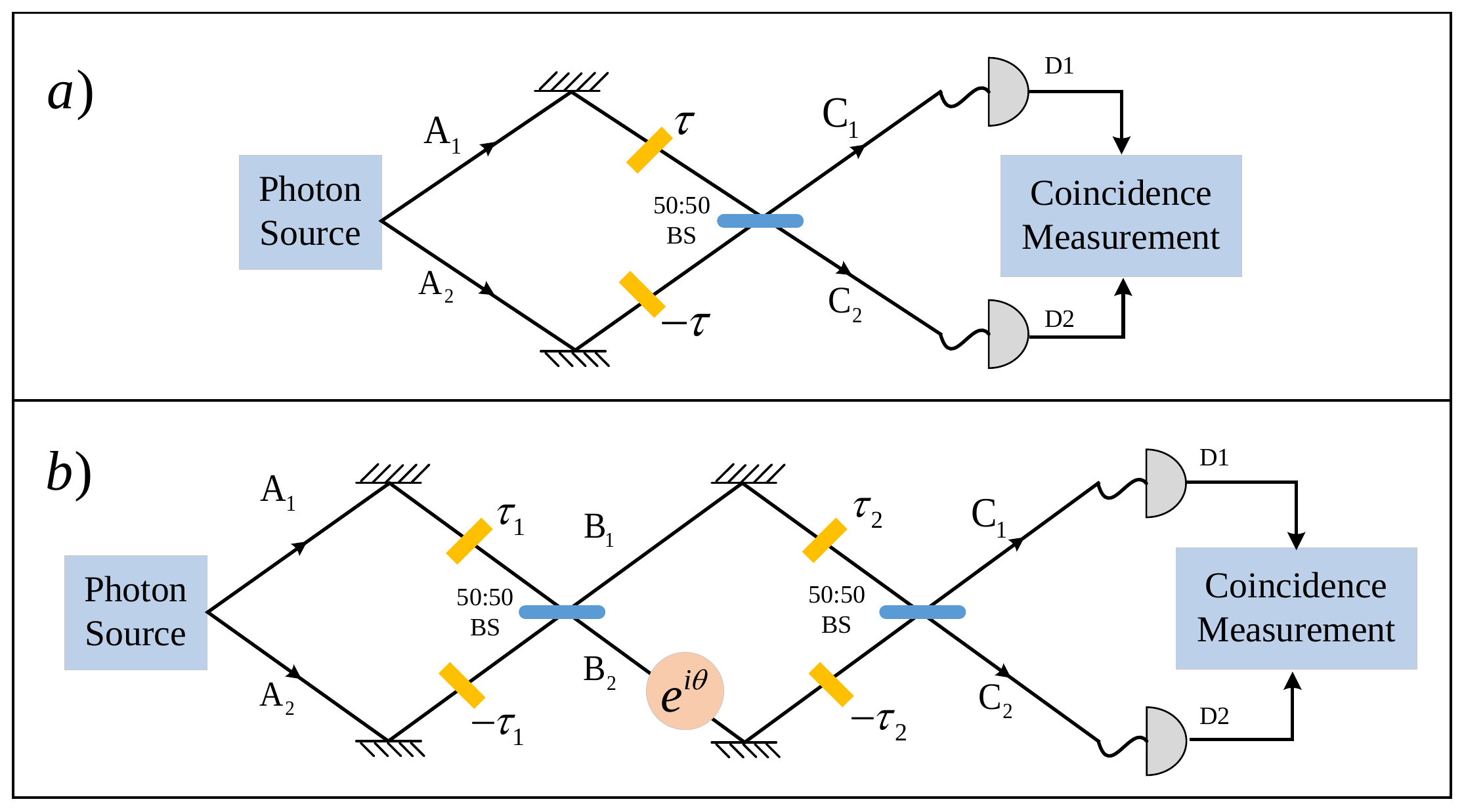} 
	\centering
	\caption{a) An outline of the HOM interferometer:  the yellow rectangles indicate 
		that  two light beams propagating on the optical paths $\mathbf{A}_1$, $\mathbf{A}_2$ experience a time delay $2\tau$. b) The sketch of our modified HOM configuration:  here there
		are two independent delays (the delay $\tau_1$  associated with the paths  $\mathbf{A}_1$, $\mathbf{A}_2$ and the delay $\tau_2$ associated with $\mathbf{B}_1$, $\mathbf{B}_2$), and 
		an achromatic waveplate~\cite{ACHRO1,ACHRO2} (pink circle in the figure), introducing a constant phase shift $e^{i \theta}$  between $\mathbf{B}_1$, $\mathbf{B}_2$.}
	\label{Fig1} 
\end{figure*}

In order to deliver our findings, we start reviewing the basic principles of the original HOM interferometer, discussing explicitly in which sense it can provide enhancements in sensing by analyzing its performances under the quantum and classical inputs. While not containing new contributions, we make an effort to presenting them in a compact form which we think is important to better appreciate the rest of the manuscript. In any case, readers who are familiar with the original HOM scheme can give a quick look at this part and moving directly to the next section where we give an explicit construction to positively achieve our main goal. Since it is obtained by concatenating the original HOM interferometer with a second 50:50 beam splitter, the proposed scheme resembles the Mach-Zhender interferometer (MZI) setup studied e.g. in  Refs.~\cite{MZ0,MZ01,MZ02,MZ03,MZ04,MZ1,MZ2}: at variance with these works however our model includes also the presence of  an achromatic wave-plate~\cite{ACHRO1,ACHRO2} which induces a frequency independent phase shift on the propagating signals, see Figure~\ref{Fig1}b). In addition, these works presented in Refs.~\cite{MZ0,MZ01,MZ02,MZ03,MZ04,MZ1,MZ2} does not exhibit the required one-to-one correspondence between the zero-coincidence event and the contemporary nullification of two delays.
Our architecture is also in part reminiscent of the proposal of Ref.~\cite{GIANNI} which, in an effort to allow for an unambiguous reconstruction for the input state of the BP source by means of the coincidence measurements, also discuss a modification of the HOM scheme capable to accommodate two independent  parameters introduced in the setup as independent delays: such scheme however requires different resources (i.e. two extra (zero-photon) ancillary modes, four beam splitters, and four detectors instead of an achromatic phase-shift, a single beam splitter, and two detectors).

As the first application of our finding we hence move to analyze the performances of the proposed setup by comparing the results obtained from the BP source with those associated with the classical input states of the light (i.e. sources with positive Glauber-Sudarshan  $P$-representation~\cite{ROY,SUD}). In this context we show that as long as the input configurations are symmetric under exchanging the input ports of the interferometer, our modified HOM setting allows one to establish a clear distinction between quantum and classical signals. Only the former being capable to deliver an exact zero for the output coincidence counts associated with the path mismatches in both the two spatial coordinates we are probing. Interestingly enough this ability in discriminating between classical and quantum light sources is not a property that holds for the original HOM setting, where symmetric coherent light pulses can also produce zero coincidences in the case of zero delay. 
We then move to consider the applications for sensing. In particular we focus on the scenario where the two delays we are probing with the modified HOM interferometer are affected by strong fluctuations (larger than the mean wavelength of the laser pump that is used to generate the BP state in the conventional optical experiments). Under this circumstance, it turns out that the effect of the achromatic phase shift which was necessary to deliver the zero coincidence condition in the ideal case, is irrelevant, its effect being washed away by the instability of the system parameters. As a matter of fact, irrespectively from the presence of the achromatic phase shift,  the possibility of observing an exact zero in the coincidence counts derived from our modified HOM scheme is lost. Nevertheless we show that two-photon input signals are still capable to provide an advantage in terms of the visibility of the measured signals compared with the classical sources. The presence of loss and dispersion is also analyzed showing that as long as they do not impact on the symmetry of the scheme, they will not affect the advantage detailed above. Next, following the original proposal by Bahder in Ref.~\cite{Bahder:2004} we also discuss how our scheme could be utilized to realize a Quantum Positioning System (QPS) which employs the coincidence counts of a single bi-photon source and two detectors to localize a collaborative target on a two-dimensional manifold. The paper finally ends with the conclusive remarks. Technical derivations are presented in the Supplementary Material online.

\section*{HOM interferometry}
In this section we review the main aspects of the HOM interferometry. As already anticipated in the introduction, readers who have familiarity with these notions can probably give a brief look to the first part where we introduce the notation, skip the rest, and move directly to the next section.
\subsection*{Notation and settings} 
 In its simplest implementation the HOM interferometer is graphically described in Figure~\ref{Fig1} a):  two (possibly correlated) input optical signals propagating along two optical paths $\mathbf{A}_1$, $\mathbf{A}_2$ characterized by a length difference $\Delta \ell$,  impinge into a 50:50 beam-splitter BS and emerge from the output ports $\mathbf{C}_1$, $\mathbf{C}_2$ where they are finally collected by the two photon-detectors $D_1$ and $D_2$.  Here coincidence photon-counts are recoded via intensity-intensity  measurements.  Following standard theory of photo-detection~\cite{MANDELBOOK} the latter  can be expressed as 
 \begin{eqnarray} \label{DEFRHOM} 
 R(\tau) = \int dt_1 \int dt_2  I(t_1,t_2)\;,
 \end{eqnarray} 
 with $I(t_1,t_2)$  the correlation function
 \begin{eqnarray} \label{INTINT}
 I(t_1,t_2):= \left\langle \hat{E}_{1}^{(-)}(t_1)\hat{E}_{2}^{(-)}(t_2)\hat{E}_{2}^{(+)}(t_2)\hat{E}_{1}^{(+)}(t_1)\right\rangle_{\text{in}}, 
 \label{Eq.19}
 \end{eqnarray} 
 where $\left\langle  \cdots \right\rangle_{\text{in}}:= \mbox{Tr}[ \cdots \hat{\rho}_{\text{in}}]$ indicates  that we are taking the expectation value with respect to  the density matrix $\hat{\rho}_{\text{in}}$ which describes the light beams entering the ports $\mathbf{A}_1$, $\mathbf{A}_2$. The  operators $\hat{E}_{j}^{(-)}(t):=\frac{1}{\sqrt{2\pi}} \int_{-\infty}^{+\infty}d\omega \hat{c}^{\dagger}_{j}(\omega) e^{i\omega t}$ and $\hat{E}_{j}^{(+)}:= (\hat{E}_{j}^{(-)})^{\dagger}$ in equation~(\ref{INTINT}) represent instead the amplitude electromagnetic field at detector $D_j$, written  in  the Heisenberg representation and formally expressed in terms of the bosonic annihilation operators $\hat{c}_j(\omega)$, describing a photon of frequency $\omega$  that emerges in the output path $\mathbf{C}_j$ and fulfilling  Canonical Commutation Rules  (CCR), e.g.  $[\hat{c}_j(\omega), \hat{c}_{j'}(\omega')]=0$, $[\hat{c}_j(\omega), \hat{c}^\dag_{j'}(\omega')]=\delta_{j,j'} \delta(\omega-\omega')$, ($\delta_{j,j'}$ and $\delta(\omega - \omega')$ representing respectively Kronecker and Dirac deltas). By direct integration equation~(\ref{DEFRHOM}) can hence be expressed in the compact form 
 \begin{eqnarray} \label{DEFRHOM1} 
 R(\tau) :=  \left\langle \hat{N}^{(c)}_1 \hat{N}^{(c)}_2 \right\rangle_{\text{in}} \;, 
 \end{eqnarray} 
 where  for $j=1,2$,  
 $\hat{N}^{(c)}_j := \int d\omega \hat{c}_j^\dag(\omega) \hat{c}_j(\omega)$ 
 is the  number operator  that counts the photons impinging on the $j$-th detector.

 An explicit characterization of the device can be obtained by assigning the input-output mapping that links the bosonic annihilation operators   $\hat{c}_j(\omega)$ with their input counterparts $\hat{a}_j(\omega)$, describing the photons of frequency $\omega$ that  propagate along the path $\mathbf{A}_j$, also  fulfilling independently CCR and used to define $\hat{\rho}_{\text{in}}$. Introducing the delay time $2\tau := \Delta \ell/c$ experienced by two incoming beams, such mapping writes
 \begin{eqnarray} 
 \hat{c}_1(\omega) 
 &=& ({\hat{a}_1(\omega) e^{- i \omega \tau} + \hat{a}_2(\omega) e^{ i \omega \tau}})/{\sqrt{2}} \;,\nonumber  \\ 
 \hat{c}_2(\omega) 
 &=& ({\hat{a}_1(\omega) e^{- i \omega \tau} - \hat{a}_2(\omega) e^{ i \omega \tau}})/{\sqrt{2}} \;, \label{SCA} 
 \end{eqnarray} 
 where the BS is balanced.
\subsection*{The dip with frequency correlated bi-photon (BP) state} 
 Let us then consider the case where the density operator $\hat{\rho}_{\text{in}}$  is represented by a frequency correlated BP pure state
 \begin{eqnarray} \label{PSI12}
 |\Psi^{(a)}_{S} \rangle : = \int d\omega \int d\omega' \Psi_S(\omega,\omega') \hat{a}_1^\dag(\omega)  \hat{a}_2^\dag(\omega') |vac\rangle\;, 
 \end{eqnarray} 
 where $|vac\rangle$ is a multi-mode vacuum state and the complex function
 $\Psi_S(\omega,\omega')$ is  an amplitude probability distribution fulfilling the normalization condition,
 $\int d\omega \int d\omega'  \left| \Psi_S(\omega,\omega')\right|^2  = \langle \Psi_{S}^{(a)}  | \Psi_{S}^{(a)}  \rangle  =1$.
 In what follows we shall assume $ \Psi_S(\omega,\omega')$ to be symmetric under exchange of the variables $\omega$ and $\omega'$, i.e. 
 \begin{eqnarray}
 \Psi_S(\omega,\omega') = \Psi_S(\omega',\omega) \;,
 \label{Eq.6}
 \end{eqnarray} 
 other choices being not capable to produce the desired results (see the subsection``Choice of the frequency spectrum function'' of Supplementary Materisl online). 
 
 \begin{figure}[!t]
 	\centering
 	\includegraphics[width=0.5\columnwidth]{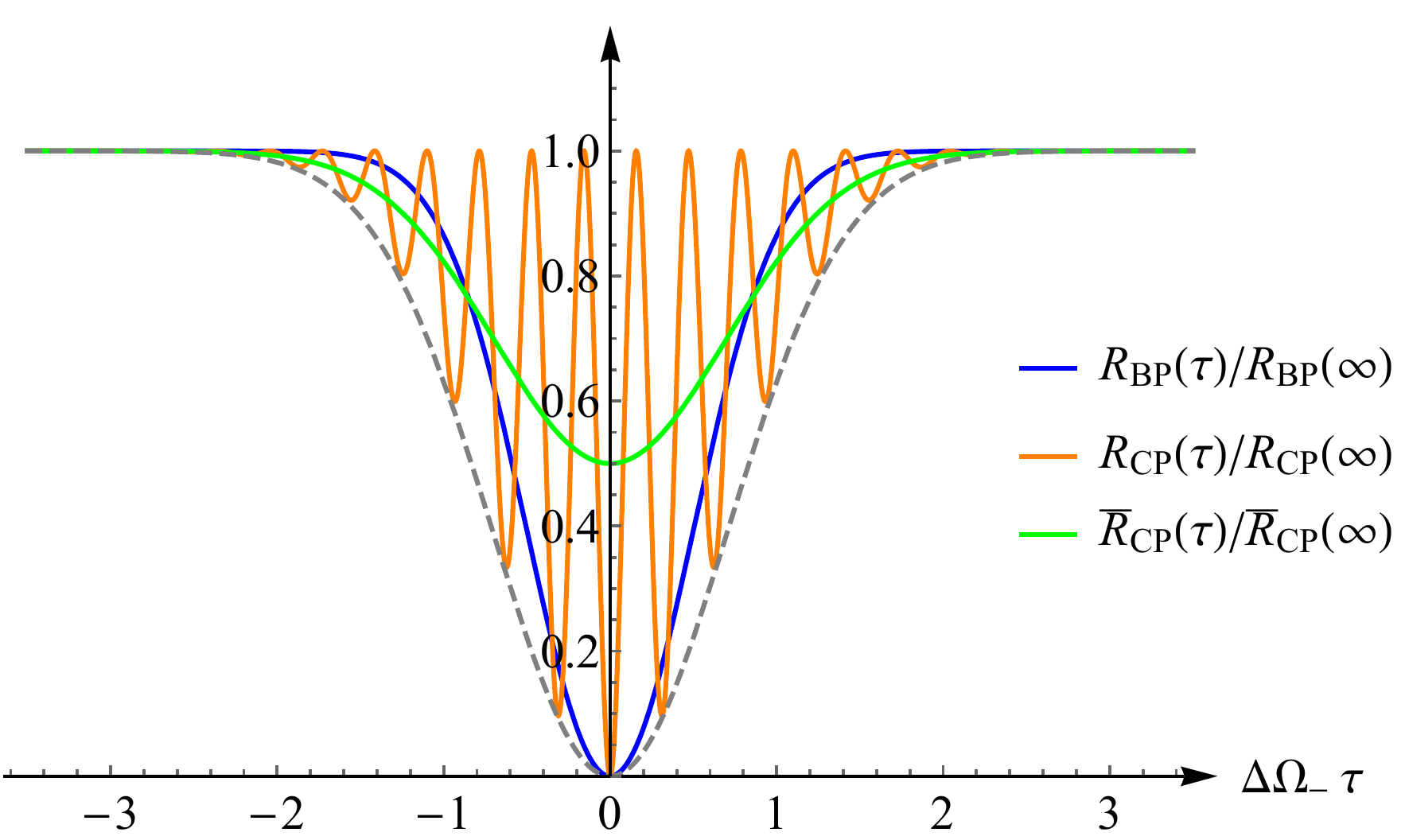}
 	\caption{Plots of HOM scheme with different input states, i.e. the BP state and the CP state. The blue curve shows the coincidence counts~(\ref{ED1}) associated with the input state $|\Psi^{(a)}_{S} \rangle$ of equation~(\ref{PSI12}) with Gaussian spectrum~(\ref{GAUS}), as a function of the delay time. It exhibits a  HOM dip at $\tau=0$ with a width which is inversely proportional to the frequency-frequency correlations spread $\Delta \Omega_-$. In addition, plot of the coincidence counts $R_{\mbox{\tiny{CP}}}(\tau)$ of equation~(\ref{COH}) (orange curve) and of its coarse grained counterpart $\overline{R}_{CP}(\tau)$~(\ref{HOMCPCORS}) (green curve), obtained from a CP state $|\alpha^{(a)}\rangle$ with local spectral width $\Delta \omega = \Delta \Omega_-/2$ (corresponding to set  $\Delta \Omega_+\ll \Delta \Omega_-$ in equation~(\ref{DEFDOMEGA})). The mean frequency $\omega_0$ was set equal to $5 \Delta \Omega_-$ for explanatory purposes: more realistic estimation would give an $\omega_0$ at least 3 order of magnitudes larger than $\Delta \Omega_-$ with more frequent oscillations for  $R_{\mbox{\tiny{CP}}}(\tau)$. All the coincidences have been renormalized with respect to their asymptotic values at $\tau\rightarrow \infty$ while the delays are reported in unit of the inverse of  $\Delta \Omega_-$. }
 	\label{S-Fig}
 \end{figure}
 
 To compute the associate value of (\ref{DEFRHOM1}) we find it convenient to express (\ref{PSI12}) in terms of the output-modes  operators by inverting equation~(\ref{SCA}), i.e. 
 \begin{eqnarray} 
 \hat{a}_1(\omega) &=& e^{ i \omega \tau} ({\hat{c}_1(\omega) + \hat{c}_2(\omega) })/{\sqrt{2}} \;,\nonumber  \\ 
 \hat{a}_2(\omega) &=& e^{ - i \omega \tau}  ({\hat{c}_1(\omega) - \hat{c}_2(\omega) })/{\sqrt{2}} \;. \label{SCAINV} 
 \end{eqnarray} 
 Accordingly we get 
 \begin{eqnarray}  
 \label{DDDD} 
 |\Psi^{(a)}_{S}\rangle =  |\Psi^{(c)}_{A}(\tau) \rangle+ |\Phi^{(c)}_-(\tau) \rangle\;,
 \end{eqnarray}  
 with  $|\Psi_{A}^{(c)} (\tau) \rangle$ and $|\Phi^{(c)}_-(\tau) \rangle$ being
 associated respectively with the events where these two photons emerge  from  opposite and identical  output ports, i.e.  
 \begin{eqnarray} 
 |\Psi^{(c)}_{A} (\tau) \rangle&: = &\int d\omega \int d\omega' \Psi^{(\tau)}_{A}(\omega,\omega') \hat{c}_1^\dag(\omega)  \hat{c}_2^\dag(\omega') |vac\rangle\;,  \label{SSS1}\\
 |\Phi^{(c)}_{-} (\tau) \rangle&: = &\frac{1}{2} \int d\omega \int d\omega' \Phi_-^{(\tau)}(\omega,\omega') \big( \hat{c}_1^\dag(\omega)  \hat{c}_1^\dag(\omega')- \hat{c}_2^\dag(\omega)  \hat{c}_2^\dag(\omega') \big) |vac\rangle\;,  \label{SS2} 
 \end{eqnarray} 
 with 
 \begin{eqnarray} 
 \Psi^{(\tau)}_{A}(\omega,\omega') &: =&  i\Psi_{S}(\omega,\omega') \sin((\omega-\omega')\tau)\;, \label{DEFPSIA} \\
 \Phi^{(\tau)}_{-}(\omega,\omega') &: =& \Psi_{S}(\omega,\omega') \cos((\omega-\omega')\tau)\;,  \label{DEFPHI-} 
 \end{eqnarray} 
 the subscript ``$A$'' of $\Psi^{(\tau)}_{A}(\omega,\omega')$  indicating that this function is anti-symmetry   (i.e. $\Psi^{(\tau)}_{A}(\omega,\omega') = - \Psi^{(\tau)}_{A}(\omega',\omega)$), while the
 subscript ``$-$'' of $\Phi^{(\tau)}_{-}(\omega,\omega')$ referring to the minus sign in the superposition (\ref{SS2}). 
 Using the fact that both $|\Psi^{(c)}_{A} (\tau) \rangle$ and $|\Phi^{(c)}_{-} (\tau) \rangle$ are eigenvectors of $\hat{N}^{(c)}_1 \hat{N}^{(c)}_2$ with eigenvalues $1$ and $0$ respectively, from (\ref{DEFRHOM1}) we get 
 \begin{eqnarray} 
 R_{\mbox{\tiny{BP}}}(\tau) 
 = \langle \Psi_{A}^{(c)}  | \Psi_{A}^{(c)}  \rangle   = \int d\omega \int d\omega' |\Psi^{(\tau)}_{A}(\omega,\omega')|^2 = \int d\nu P_S(\nu)  \sin^2(\nu \tau)\;,  \label{HOM111}
 \end{eqnarray} 
 where  
 $P_S(\nu) 
 := \int d \nu'  \big| \Psi_S(\nu'+ \nu/2,\nu'-\nu/2)\big|^2$
 is the probability distribution associated with the frequency difference for two photons in the BP state. For the sources available in the lab, (\ref{HOM111}) exhibits the typical HOM dip, i.e. a unique minimum value at $\tau=0$ with a width that is inversely proportional to the spread of  $P_S(\nu)$. 
 
 As a concrete example let us consider the paradigmatic case of a   Gaussian two-mode spectrum function 
 \begin{eqnarray}\label{GAUS} 
 |\Psi_S(\omega,\omega')|^2 =
 \tfrac{1}{\sqrt{2\pi} \Delta \Omega_+}  \exp[ - \tfrac{(\omega + \omega' - 2 \omega_0)^2}{8 \Delta^2 \Omega_+} ]\times  \tfrac{1}{\sqrt{2\pi} \Delta \Omega_-}  \exp[ - \tfrac{(\omega-\omega')^2}{2 \Delta^2 \Omega_-} ] \;, 
 \end{eqnarray} 
 which locally assigns to each  photon an average  frequency $\omega_0$ and a spread 
 \begin{eqnarray} \label{DEFDOMEGA}
 \Delta \omega: =\sqrt{ \Delta \Omega_-^2 + 4 \Delta \Omega_+^2}/2\;. 
 \end{eqnarray} 
 Under this assumption equation~(\ref{HOM111}) can be explicitly computed yielding
 \begin{eqnarray} \label{ED1} 
 R_{\mbox{\tiny{BP}}}(\tau) =(1- e^{-2\; \Delta^2 \Omega_- \tau^2})/2 \;,
 \end{eqnarray} 
with a minimum that exhibits a width proportional to $1/\Delta \Omega_-$ and a visibility (ratio $V_{\mbox{\tiny{BP}}}:=[R_{\mbox{\tiny{BP}}}(\infty)-R_{\mbox{\tiny{BP}}}(0)]/R_{\mbox{\tiny{BP}}}(\infty)$ between the depth of the dip and the height of the plateau) of $100\%$, see the blue curve of Figure~\ref{S-Fig}. This special feature facilitates the HOM interferometer as a tool for the phase drifts or displacements measurements: indeed, writing the path length difference $\Delta \ell$ of  $\mathbf{A}_1$, $\mathbf{A}_2$  as the sum of two contributions $\Delta \ell=\Delta \ell^{(0)}+  x$, the first (i.e. $\Delta \ell^{(0)}$) fixed and unknown, the second (i.e. $x$) controllable by the experimentalist (e.g. realized by inserting a crystal of variable thickness or refractive index), we can use (\ref{ED1}) to recover the value of former by simply pinpointing the zero of the function $f_{\mbox{\tiny{BP}}}(x):=R_{\mbox{\tiny{BP}}}(\tau =\frac{\Delta \ell^{(0)} + x}{2c})$  with a resulting accuracy which is proportional~\cite{SENSOR1} to the inverse of the bi-photon spectral width $c/\Delta \Omega_-$.

 Notice that equation~(\ref{ED1}) does not depend neither from the average frequency $\omega_0$ of the local signals, nor from the parameter $\Delta \Omega_+$ which gauges the spread of sum of the photons frequencies. This last observation is particularly relevant as by varying the ratio between $\Delta \Omega_+$ and $\Delta \Omega_-$, equation~(\ref{GAUS}) allows us to analyze a large variety of two-photon input states  $|\Psi_S^{(a)}\rangle$  exhibiting different degrees  of two-photon entanglement. For instance, in the limit $\Delta \Omega_+ \ll \Delta \Omega_-$, the distribution (\ref{GAUS}) approaches the frequency entangled two-photon state emerging from an ideal Spontaneous Parametric Down Conversion (SPDC) source pumped with a laser of mean frequency $\omega_p=2 \omega_0$, see e.g.~\cite{SPDC,SPDC1,SPDC2} and references therein; for $\Delta \Omega_+=\Delta\Omega_-/2$ instead  the functional dependence of $|\Psi_S(\omega,\omega')|^2$ upon $\omega$ and $\omega'$ factorizes, describing the case of two uncorrelated (unentangled) single-photon packets; finally for $\Delta \Omega_+ \gtrapprox \Delta \Omega_-$, $|\Psi_S(\omega,\omega')|^2$ exhibits a strongly correlated behavior, this time mimicking the properties of an entangled DB state emitted by a difference-beam source~\cite{TWIN}. According to equation~(\ref{ED1}) all these states exhibit  the same $\tau$ dependence of the coincidence counts, making it clear that the  peculiar features of the HOM dip cannot be linked  to the presence of entanglement in the input state: at most, since the presence of asymmetries in the BP spectrum prohibits the formation of the dip (see Ref.~\cite{SPDC2} and the subsection ``Choice of the frequency spectrum function'' of Supplementary Material online), one can claim that it is the symmetric character (indistinguishability) of the two-photon wave-packet $\Psi_S(\omega,\omega')$ that does matter.
 
\subsubsection*{The dip with coherent pulses (CP) state}
If the HOM dip is not a consequence of entanglement, in which sense it can be considered a genuine evidence of quantum interference? To answer this question one should compare equation~(\ref{ED1}) with the coincidence counts obtained when injecting into the interferometer two classical light signals one propagating along $\mathbf{A}_1$ and the other along  $\mathbf{A}_2$ and exhibiting the analogous spectral properties of the two-photon input~(\ref{PSI12}). A simple but informative way to model this situation is obtained by replacing $|\Psi^{(a)}_{S} \rangle$ with a symmetric multi-mode Coherent Pulses (CP) state $|\alpha^{(a)}\rangle$ obeying the constraints 
\begin{eqnarray} \label{COHERENT}
\hat{a}_j (\omega)  |\alpha^{(a)}\rangle= \alpha(\omega) |\alpha^{(a)}\rangle\;, \quad \forall j=1,2\;,
\end{eqnarray}
with the complex function  $\alpha(\omega)$ being the associated photon amplitude distribution at frequency $\omega$.Via  equation~(\ref{SCA}) this implies
\begin{eqnarray} 
\hat{c}_1 (\omega)  |\alpha^{(a)}\rangle&=& \sqrt{2} \cos(\omega\tau) \alpha(\omega)  |\alpha^{(a)}\rangle\;,\label{COHERENT11}\\ 
\hat{c}_2 (\omega)  |\alpha^{(a)}\rangle&=& -i \sqrt{2} \sin(\omega\tau) \alpha(\omega)  |\alpha^{(a)}\rangle\;.\label{COHERENT12}
\end{eqnarray}
Replacing this into  equation~(\ref{DEFRHOM1}) and making use of the CCR relations, a simple algebra allows us to write
\begin{eqnarray} \label{AAA}  
R_{\mbox{\tiny{CP}}}(\tau) ={A^2}\Big[ 1 -\left( \int d\omega P_\alpha(\omega)  \cos(2\omega \tau)\right)^2\Big]\;, 
\end{eqnarray} 
where $A:= \int d\omega  \left| \alpha(\omega)\right|^2$ is the total photon intensity at each of the input ports, and where $P_\alpha(\omega) := |\alpha (\omega)|^2/A$ is the probability distribution describing the frequency spectrum of the corresponding beam. For a fair comparison with equation~(\ref{ED1}), we assume 
$P_\alpha(\omega)$ to be normally distributed with the same average value $\omega_0$ and spread $\Delta \omega$ adopted for the local properties of the bi-photon state $|\Psi^{(a)}_{S} \rangle$,
obtaining 
\begin{eqnarray} \label{COH}
R_{\mbox{\tiny{CP}}}(\tau) = {A^2}(  1- \cos^2(2\omega_0\tau) e^{-4\; \Delta^2 \omega \tau^2} \big)\;.
\end{eqnarray} 
Notice that as in the case of equation~(\ref{ED1}), this expression still reaches zero if and only if  $\tau=0$ (as shown in the subsection ``Optimal CP states'' of Supplementary Material online, $|\alpha^{(a)}\rangle$ is the only coherent pulse state which admits $\tau=0$ as zero coincidence point).
Yet due to the presence of $\omega_0$ in (\ref{COH})  such minimum comes along with fast oscillations that introduce a series of extra local minima that, while observable in the controlled experimental settings, will tend to reduce the effective visibility of the dip in realistic scenarios. This phenomenon is shown in Figure~{\ref{S-Fig}} depicted by the orange curve. Indeed, assuming the delay time $\tau$ to be affected by fluctuations of intensity $T$ larger than $1/\omega_0$ but smaller than $1/\Delta \omega$, the measured value of coincidence counts will being not longer provided by $R(\tau)$ but instead by its coarse grained counterpart
\begin{eqnarray}  \label{CORS}
\overline{R}(\tau): = \frac{1}{T}\int_{\tau-T/2}^{\tau+T/2} d\tau' R(\tau')\;,
\end{eqnarray} 
(for optical system setting, the typical values would be $\omega_0\simeq10^{16} {\rm Hz}$ and $\Delta \Omega_-\simeq 10^{12} {\rm Hz}$). For the BP state such averaging does not introduce significant modifications, i.e. $\overline{R}_{\mbox{\tiny{BP}}}(\tau)\simeq {R_{\mbox{\tiny{BP}}}(\tau)}$. For the CP state instead this implies 
\begin{eqnarray} \label{HOMCPCORS}
\overline{R}_{\mbox{\tiny{CP}}}(\tau)\simeq  {A^2} \big(1- \tfrac{1}{2} e^{-4\; \Delta^2 \omega \tau^2}\big)\;,
\end{eqnarray}
which we report in Figure~\ref{S-Fig} with a green curve. This function no longer nullifies at $\tau=0$ but admits this value as a unique minimum with a width proportional to $1/\Delta \omega$ which, due to~(\ref{DEFDOMEGA}), typically is comparable with the width of  (\ref{ED1}), at least for $\Delta \Omega_+$ larger than or of the same order of $1/\Delta \Omega_-$. The main difference however is the drastic reduction in visibility which pass from the $100\%$ of the BP state to the $50\%$ of the CP state. This main difference exactly proves that photons with BP state as the inputs is reasonable in the traditional HOM scheme. To summarise, the existence of a unique zero of the function $R(\tau)$ is not only peculiar for the entangled photon state, which still occurs for the classical lights. Instead quantum advantages can be identified in the enhancement of visibility of such minimum, the carrier frequency of the local signals ($2\omega_0$) induces an effective coarse graining of the measured signals in realistic scenarios.

\begin{figure*}
	\centering
	\includegraphics[width=1\columnwidth]{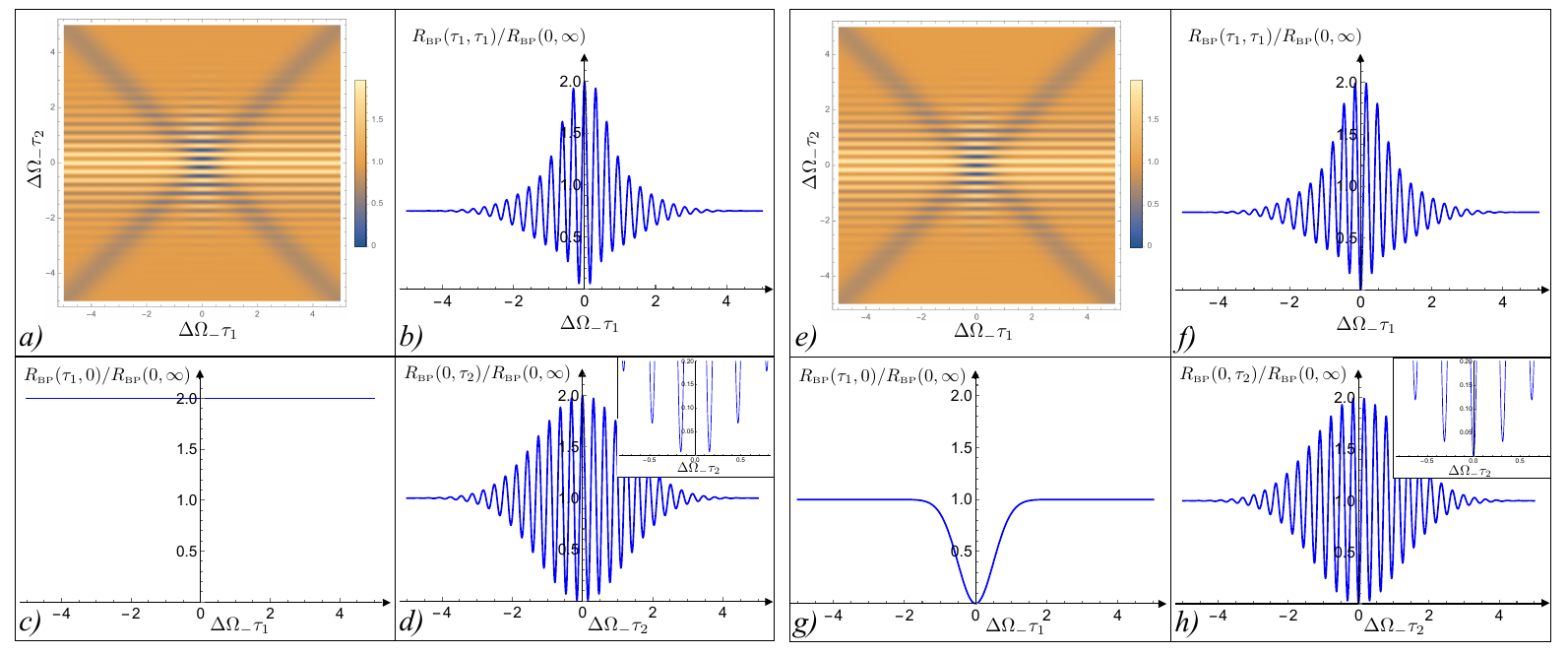} 
	\caption{{Coincidence counts ${R}_{\mbox{\tiny{BP}}}(\tau_1,\tau_2)$ of the BP state according to equation~(\ref{ED1222}): 
			the left panels  from a) to d)  refer to the case of  $\theta=0$ (no achromatic phase shift), formally equivalent to the configurations analyzed in Refs.~\cite{MZ0,MZ01,MZ02,MZ03,MZ04,MZ1,MZ2}; the right panels from e) to h) instead refer to the case of $\theta=\pi/2$ (a quarter-wave plate contributes the achromatic phase shift). 
			Panels a) and e): contour plots of  ${R}_{\mbox{\tiny{BP}}}(\tau_1,\tau_2)$;
			Panels b) and f)  plots of the function ${R}_{\mbox{\tiny{BP}}}(\tau_1,\tau_1)$; 
			Panels c) and g) plots of ${R}_{\mbox{\tiny{BP}}}(\tau_1,0)$; 
			Panels d) and h) finally report ${R}_{\mbox{\tiny{BP}}}(0,\tau_2)$. 	
			In all the figures the delays times have been rescaled by the inverse of the width $\Delta \Omega_-$  of the bi-photon spectrum, and the rates have been rescaled by the plateau value ${R}_{\mbox{\tiny{BP}}}(0,\infty)=1/2$. The mean frequency $\omega_0$ was set equal to $5 \Delta \Omega_-$ and $\Delta \Omega_+=\Delta \Omega_-/5$. Notice that  ${R}_{\mbox{\tiny{BP}}}(\tau_1,\tau_1)$ is strictly positive  for $\theta=0$, while for $\theta=\pi/2$ it reaches zero  only for $\tau_1=\tau_2=0$. }}
	\label{BP_CHANGE_TAU}
\end{figure*}

\section*{Modified HOM configuration}
In this section our modified HOM configuration is presented to show how it can be used for identifying two independent delays with a unique zero of  the output coincidence counts in the ideal case. 
As we shall see, the key ingredient to achieve this goal is the presence of the achromatic phase shifter placed along one of the paths that connects the two beam splitters.
We remind that unlike standard wave-plates,  this element provides a constant phase shift independent with the wavelength of the incoming light, typically achieved by using two different birefringent crystalline materials balancing the relative shift in retardation over the wavelength range~\cite{ACHRO1,ACHRO2}.

\subsection*{Settings}
The scheme we consider is still based on recording coincidences in the photon-counts at the output of a two-ports interferometer. Yet at variance with the original HOM setting it includes two 50:50 BSs and  two independent delays elements associated with the parameters $\Delta \ell_1$ and $\Delta \ell_2$ which are located before the first and second BS respectively, see Figure~\ref{Fig1} b). Most importantly, we also assume the presence of an achromatic plate ~\cite{ACHRO1,ACHRO2} operating between the two BSs. For $j=1,2$ we shall indicate with $\mathbf{A}_j$, $\mathbf{B}_j$, and $\mathbf{C}_j$ the paths composing the interferometers and with $\hat{a}_j(\omega)$, $\hat{b}_j(\omega)$, and $\hat{c}_j(\omega)$ the associated annihilation operators, each fulfilling CCR rules and being linked by proper scattering unitary transformations. In particular, the relation connecting  $\hat{a}_j(\omega)$ and $\hat{b}_j(\omega)$ is the same as the one we introduced for the HOM configuration, i.e. 
\begin{eqnarray} 
\hat{b}_1(\omega) &=&  \tfrac{\hat{a}_1(\omega) e^{- i \omega \tau_1} + \hat{a}_2(\omega) e^{ i \omega \tau_1}}{\sqrt{2}} \;,\nonumber  \\ 
\hat{b}_2(\omega) &=&   \tfrac{\hat{a}_1(\omega) e^{- i \omega \tau_1} - \hat{a}_2(\omega) e^{ i \omega \tau_1}}{\sqrt{2}} \;. \label{SCA121} 
\end{eqnarray} 
where as in the previous section we introduce the variable
\begin{eqnarray}
\tau_j : = \Delta \ell_j/(2c)\;,
\end{eqnarray}
to represent the temporal shift associated with the path length difference $\Delta \ell_j$.
Similarly for $\hat{b}_j(\omega)$ and $\hat{c}_j(\omega)$ we have 
\begin{eqnarray} 
\hat{c}_1(\omega) &=&  \tfrac{\hat{b}_1(\omega) e^{- i \omega \tau_2} + e^{i\theta} \hat{b}_2(\omega) e^{ i \omega \tau_2}}{\sqrt{2}} \;,\nonumber  \\ 
\hat{c}_2(\omega) &=&  \tfrac{\hat{b}_1(\omega) e^{- i \omega \tau_2} - e^{i\theta} \hat{b}_2(\omega) e^{ i \omega \tau_2}}{\sqrt{2}} \;, \label{SCA122} 
\end{eqnarray} 
where we included  the achromatic waveplate contribution~\cite{ACHRO1,ACHRO2}  which adds a constant phase $\theta$ to the photon propagating along the path $\bf{B}_2$  irrespectively from its frequency. The resulting input-output relations are hence obtained by replacing (\ref{SCA121}) into (\ref{SCA122}), and  upon some simple algebra they can be expressed in the following compact form 
\begin{eqnarray} \label{SCA1} 
\hat{c}_1(\omega)&=& \cos(\omega \tau_2+\theta/2)e^{-i(\omega\tau_1-\theta/2)}  a_1(\omega)-  i\sin(\omega \tau_2+\theta/2)e^{i(\omega\tau_1+\theta/2)} a_2(\omega)\;,  \nonumber  \\
\hat{c}_2(\omega)&=& - i \sin(\omega \tau_2+\theta/2)e^{-i(\omega\tau_1-\theta/2)}  a_1(\omega)+ \cos(\omega \tau_2+\theta/2)e^{i(\omega\tau_1+\theta/2)} a_2(\omega)\;.
\end{eqnarray} 
\subsection*{Interference performances}
From (\ref{SCA1}) it follows that the BP input state (\ref{PSI12}) expressed in terms of the output operators $\hat{c}_j(\omega)$ of the modified HOM scheme becomes
\begin{eqnarray} 
\label{DDDD1} 
|\Psi^{(a)}_{S}\rangle =  |\Psi^{(c)}(\tau_1,\tau_2) \rangle+ |\Phi^{(c)}(\tau_1,\tau_2) \rangle\;,
\end{eqnarray}  
where $|\Psi^{(c)}(\tau_1,\tau_2) \rangle$ represents a component with one photon per each output ports and $|\Phi^{(c)}(\tau_1,\tau_2) \rangle$ the one where instead both photons emerge from the same port, either ${\bf C}_1$ or ${\bf C}_2$. In particular $|\Psi^{(c)}(\tau_1,\tau_2) \rangle$ is characterized by two components  having, respectively, anti-symmetric and symmetric two-photon amplitude spectra, i.e. 
\begin{eqnarray} 
\label{DDDD11} 
|\Psi^{(c)}(\tau_1,\tau_2) \rangle &= & |\Psi^{(c)}_A(\tau_1,\tau_2) \rangle+ |\Psi^{(c)}_S(\tau_1,\tau_2) \rangle\;, \\ 
|\Psi^{(c)}_{A,S} (\tau_1,\tau_2) \rangle&: =&  \int d\omega \int d\omega' \Psi^{(\tau_1,\tau_2)}_{A,S}(\omega,\omega') \hat{c}_1^\dag(\omega)  \hat{c}_2^\dag(\omega') |vac\rangle\;,   
\nonumber 
\end{eqnarray} 
with 
\begin{eqnarray} 
\Psi^{(\tau_1,\tau_2)}_{A}(\omega,\omega') &: =& -i e^{i\theta}\Psi_{S}(\omega,\omega') \sin((\omega-\omega')\tau_1) \cos((\omega-\omega')\tau_2)\;,   \\
\Psi^{(\tau_1,\tau_2)}_{S}(\omega,\omega') &: =&e^{i\theta} \Psi_{S}(\omega,\omega') \cos((\omega-\omega')\tau_1)    \cos((\omega+\omega')\tau_2+\theta)\;,
\end{eqnarray} 
(an analogous decomposition  for $|\Phi^{(c)}(\tau_1,\tau_2) \rangle$ is reported in the subsection ``Decomposition of the output state'' of Supplementary Material online). From the above identities we can compute the coincidence counts at $D_1$ and $D_2$, which once more is provided by the expression (\ref{DEFRHOM1}). As in the standard HOM setting it is helpful to observe that the vectors $| \Psi^{(c)} (\tau) \rangle$ and $| \Phi^{(c)} (\tau) \rangle$ are eigenvectors of $\hat{N}^{(c)}_1 \hat{N}^{(c)}_2$ with eigenvalues $1$ and $0$ respectively. Accordingly we get 
\begin{eqnarray} \label{HOM1112}
R_{\mbox{\tiny{BP}}}(\tau_1,\tau_2) &=&\langle \Psi^{(c)}(\tau_1,\tau_2)  | \Psi^{(c)}(\tau_1,\tau_2)  \rangle =\langle \Psi_{A}^{(c)} (\tau_1,\tau_2) | \Psi_{A}^{(c)}(\tau_1,\tau_2)  \rangle   +\langle \Psi_{S}^{(c)}(\tau_1,\tau_2)  | \Psi_{S}^{(c)}(\tau_1,\tau_2)  \rangle \;, \nonumber\\
&=& \int d\omega \int d\omega' |\Psi_{S}(\omega,\omega')|^2 \Big( \sin^2((\omega-\omega')\tau_1)  \cos^2((\omega-\omega')\tau_2) + \cos^2((\omega-\omega')\tau_1)  \cos^2((\omega+\omega')\tau_2+\theta)\Big) \;, \nonumber \\
\end{eqnarray} 
where in the second identity we used the fact that $|\Psi^{(c)}_A(\tau_1,\tau_2)\rangle$ and $|\Psi^{(c)}_S(\tau_1,\tau_2)\rangle$ are orthogonal, i.e. $\langle \Psi_{A}^{(c)}(\tau_1,\tau_2)|\Psi_{S}^{(c)}(\tau_1,\tau_2)\rangle=0$. 
Focusing on the paradigmatic case of the Gaussian spectrum of equation~(\ref{GAUS}) this  yields
\begin{eqnarray} \label{ED1222}
R_{\mbox{\tiny{BP}}}(\tau_1,\tau_2)=\frac{1}{8}\Big[4+2e^{-2\Delta^2 \Omega_{-} \tau_2^2} -e^{-2\Delta^2 \Omega_{-}(\tau_1+\tau_2)^2}	-e^{-2\Delta^2 \Omega_{-}(\tau_1-\tau_2)^2} +2 e^{-8 \Delta^2 \Omega_{+} \tau_2^2}(1+e^{-2\Delta^2 \Omega_{-} \tau_1^2})\cos(4\tau_2 \omega_0+2\theta)\Big]\;. 
\end{eqnarray}
Equation~(\ref{ED1222}) it is the formal counterpart of the two-dimensional coincidence count quantities used in Ref.~\cite{GIANNI} for the spectral reconstruction of the BP state, and by setting the achromatic phase shift $\theta=0$ (which is effectively equivalent to remove such element from the setup),  reproduces the interference fringes observed e.g. in Refs.~\cite{MZ04,MZ1,MZ2,MZ02,MZ03}.

\begin{figure*}
	\centering
	\includegraphics[width=0.5\columnwidth]{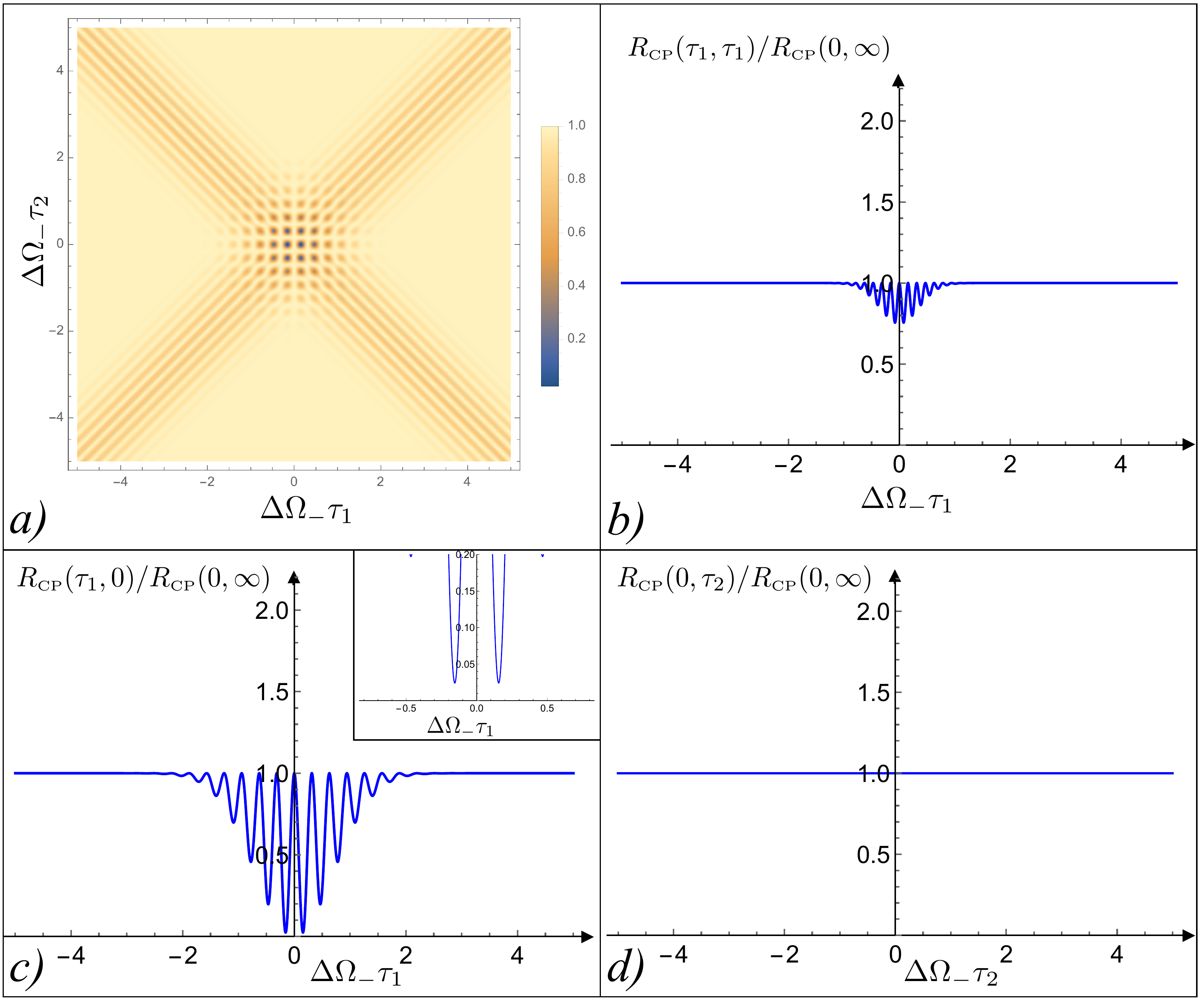} 
	\caption{{Coincidence rate ${R}_{\mbox{\tiny{CP}}}(\tau_1,\tau_2)$ of the CP state
			according to equation~(\ref{CP1asdfad}) for  $\theta=\pi/2$ . Panel a):  contour  plot of ${R}_{\mbox{\tiny{CP}}}(\tau_1,\tau_2)$;
			Panels b), c) and d): plots of function ${R}_{\mbox{\tiny{CP}}}(\tau_1,\tau_1)$, ${R}_{\mbox{\tiny{CP}}}(\tau_1,0)$ and ${R}_{\mbox{\tiny{CP}}}(0,\tau_2)$.       
			In all the figures the delays times have been rescaled by the inverse of the width $\Delta \Omega_-$  of the bi-photon spectrum, and the rates have been rescaled by the plateau value ${R}_{\mbox{\tiny{CP}}}(0,\infty)=A^2$.  Notice that differently from the BP case reported in Figure~\ref{BP_CHANGE_TAU}, this function never reaches zero: in particular at the origin of the coordinate axis it reaches the value $1$ (this result being independent from the value of $\theta$). 
			The mean frequency $\omega_0$ was set equal to $5 \Delta \Omega_-$ and $\Delta \Omega_+=\Delta \Omega_-/5$ so that $\Delta \omega\simeq \Delta \Omega_-/2$ -- see equation~(\ref{DEFDOMEGA}).
	}}
	\label{FIGURENEWCP1}
\end{figure*}

A close inspection  reveals that for arbitrary values of the achromatic phase shift $\theta$, (\ref{ED1222}) is strictly positive, see Figure~\ref{BP_CHANGE_TAU}. The only exception occurs when we set the achromatic phase shift into the quarter-wave plate configuration ($\theta=\pi/2$), which gives  
\begin{eqnarray}\label{SPECIAL} 
R_{\mbox{\tiny{BP}}}(0,0)\Big\rvert_{\theta=\pi/2} = \frac{1+\cos(2\theta)}{2}
\Big\rvert_{\theta=\pi/2}=0\;,
\end{eqnarray} 
making the origin of coordinate axis an absolute minimum of the coincidence counts. From this observation it clearly emerges that setting $\theta=\pi/2$ is a fundamental ingredient in our construction: imposing it  we can ensure that the differences of paths  $\mathbf{A}_1\mathbf{A}_2$ and $\mathbf{B}_1\mathbf{B}_2$ of Figure~\ref{Fig1} b)  can be simultaneously identified by the unique zero of $R_{\mbox{\tiny{BP}}}(\tau_1,\tau_2)$, mimicking the behavior observed in the original HOM scheme where the unique zero of the coincidence counts is associated with the difference of  paths $\mathbf{A}_1$ and $\mathbf{A}_2$ of Figure~\ref{Fig1} a). An intuitive explanation of why $\theta=\pi/2$ is such an exceptional choice for the achromatic phase shift element is presented in the subsection ``The role of the achromatic phase shift'' of Supplementary Material online by studying the evolution of the BP state along the interferometer:
basically, by setting $\theta=\pi/2$ we force an exact swap between the bi-photon state with +superposition and the one with -superposition, and both of these two states are symmetric and correspond to the event where two photons emerge from the first BS along the same path (either $\mathbf{B}_1$ or $\mathbf{B}_2$). This statement is also presented in the previous paragraph of  Eq.(B31) of the Supplementary Material online. This transformation, along with the condition to have both $\tau_1$ and $\tau_2$ exactly equal to zero, is the very reason why ultimately zero-coincidence can be recorded at the output of our modified HOM scheme.

\section*{BP vs CP performances comparison in the modified HOM scheme}
The behaviour of  $R_{\mbox{\tiny{BP}}}(\tau_1,\tau_2)$ reported in the previous section is better appreciated when comparing it with the output coincidence counts obtained when feeding the modified HOM interferometer with the classical light input described by the CP state~(\ref{COHERENT}). 
Under the Gaussian assumption on the frequency spectrum, equation~(\ref{ED1222}) gets replaced by 
\begin{eqnarray} 
R_{\mbox{\tiny{CP}}}(\tau_1,\tau_2)  =A^2\left\lbrace 1-\frac{1}{4}\Big[\cos(\theta+2\omega_0(\tau_2+\tau_1)) e^{-2\Delta^2\omega(\tau_1+\tau_2)^2}-\cos(\theta+2\omega_0(\tau_2-\tau_1))e^{-2\Delta^2\omega(\tau_1-\tau_2)^2}\Big]^2\right\rbrace \;, 
\label{CP1asdfad} 
\end{eqnarray}
which we plot in Figure~\ref{FIGURENEWCP1} (the parameters $A$, $\omega_0$, and $\Delta \omega$ being the same as in equation~(\ref{COH})). One may notice that in this case, irrespectively from the selected value of  the achromatic shift $\theta$,  $R_{\mbox{\tiny{CP}}}(\tau_1,\tau_2)$ never reaches zero: in particular at the origin of the  $(\tau_1,\tau_2)$ plane we have 
\begin{eqnarray} \label{EEE} 
R_{\mbox{\tiny{CP}}}(0,0) = A^2\;, 
\end{eqnarray} 
for all $\theta$s. We can hence conclude that the presence of a zero coincidence point obtainable in our setup for $\theta=\pi/2$, is a peculiar feature of the BP state which cannot be replicated by iso-spectral classical pulses. It is worth stressing that this effect was not present in the original HOM setting where the unicity of the zero in the coincidence count was not a peculiarity  of the BP state, as it was also obtainable by employing the CP state as input, see Figure~\ref{S-Fig}.

More generally, the impossibility of getting a zero value for the coincidence counts associated with $\tau_1=\tau_2=0$ can be shown to apply to  any input state of light  $\hat{\rho}_{\mbox{\tiny{CL-S}}}$ which, irrespectively from its spectral properties,  is classical in the sense of possessing a positive Glauber-Sudarshan  $P$-representation~\cite{ROY,SUD}, and which is symmetric under exchange of the input ports of the interferometer. $\hat{\rho}_{\mbox{\tiny{CL-S}}}$ can be expressed as a convex mixture of  the multi-mode coherent states $|\alpha^{(a)}\rangle$ obeying the symmetric constraint of equation~(\ref{COHERENT}), i.e.
$\hat{\rho}_{\mbox{\tiny{CL-S}}}:= \int d^2 \mu(\alpha) |\alpha^{(a)}\rangle \langle \alpha^{(a)}|$, with
$d^2 \mu(\alpha)$ being a positive measure on the space of the amplitudes $\alpha(\omega)$. Accordingly we get
\begin{eqnarray} \label{EEE1} 
R_{{\mbox{\tiny{CL-S}}}}(\tau_1,\tau_2) =  \int d^2 \mu(\alpha)  \left[ \left( \int d\omega |\alpha(\omega)|^2 \right)^2- \left( \int d\omega |\alpha(\omega)|^2 \sin(2 \omega \tau_2 +\theta) \sin(2\omega \tau_1) \right)^2\right] \;, 
\end{eqnarray} 
which for $\tau_1=\tau_2=0$ yields indeed
\begin{eqnarray} 
R_{{\mbox{\tiny{CL-S}}}}(0,0) =  \int d^2 \mu(\alpha)  \left( \int d\omega |\alpha(\omega)|^2 \right)^2>0\;.
\end{eqnarray} 

\section*{Sensing applications} 
This section is devoted to study whether one can use the sensitivity of BP source propagating into the modified HOM scheme for multi-parameter sensing applications. As the before analyses in the original HOM scheme, we start by briefly discussing the case where the time delays $\tau_1$ and $\tau_2$ are stable (see the subsection ``Case 1''). 
Secondly we focus on the scenario where the scheme is affected by the external noise sources. Specifically in the subsection ``Case 2'', we address  the situation in which the parameters we wish to determine are effected by the uncontrollable fluctuations during the acquisition time of the coincidence counts.
Thirdly, in the subsection ``Case 3'' we also include the presence of loss and dispersion along the optical paths composing the interferometer. 
It is worth stressing that in the last two cases the achromatic phase shifter plays no role since its effect is washed away by the noise: accordingly the analyses we present would produce the same result when applied to the Mach-Zehnder schemes of Refs.~\cite{MZ0,MZ01,MZ02,MZ03,MZ04,MZ1,MZ2} that differ from ours by having no the achromatic phase shifter. 
This section finally ends with a possible application in the context of QPS. 

We anticipate that in our presentation we shall explicitly avoid to use the technical tool borrowed from quantum metrology~\cite{METROLOGY}, instead of the figure of merit -- visibility that has been discussed more in the conventional HOM scheme. This approach, while not being conclusive to certify the quality of our scheme, has the merit of simplifying the analyses and keeping the discussions at an intuitive level. 

\subsection*{Case 1: ideal setting}
Having set the achromatic phase shift equals to the optimal quarter-wave plate configuration $\theta=\pi/2$, consider the case where, analogously to what we have done when reviewing the original HOM scheme, the delays $\Delta \ell_1$ and $\Delta \ell_2$  of Fig.~\ref{Fig1} b), are composed by two (independent) unknown contributions $\Delta \ell_1^{(0)}$ and $\Delta \ell_2^{(0)}$ plus the controllable terms $x_1$ and $x_2$ one can modulate to compensate the former, i.e. $\Delta \ell_j=  \Delta \ell_j^{(0)} + x_j$. 
If all these parameters are sufficiently stable with fluctuations smaller than the carrier wavelength $\lambda_0=2\pi c/\omega_0$ (an assumption that we shall expound in the next subsection), then clearly one can use (\ref{ED1222}) to identify $\Delta \ell_1^{(0)}$ and $\Delta \ell_2^{(0)}$ by spanning the $(x_1, x_2)$ plane to pinpoint the unique zero of $R_{\mbox{\tiny{BP}}}(\tau_1= \frac{\Delta \ell_1^{(0)}+ x_1}{2c},\tau_2= \frac{\Delta \ell_2^{(0)} + x_2}{2c})$. As in the original HOM setting~\cite{SENSOR1} the resulting accuracy is associated with the width of the minimum which for $\Delta \ell_1^{(0)}$ is the order of $c/\Delta \Omega_-$, and for $\Delta \ell_2^{(0)}$ is the order of the pump carrier wavelength $\lambda_0$ -- see panels g) and h) of Figure~\ref{BP_CHANGE_TAU}.

\subsection*{Case 2:  fluctuating parameters}
\begin{figure*}
	\includegraphics[width=0.5\columnwidth]{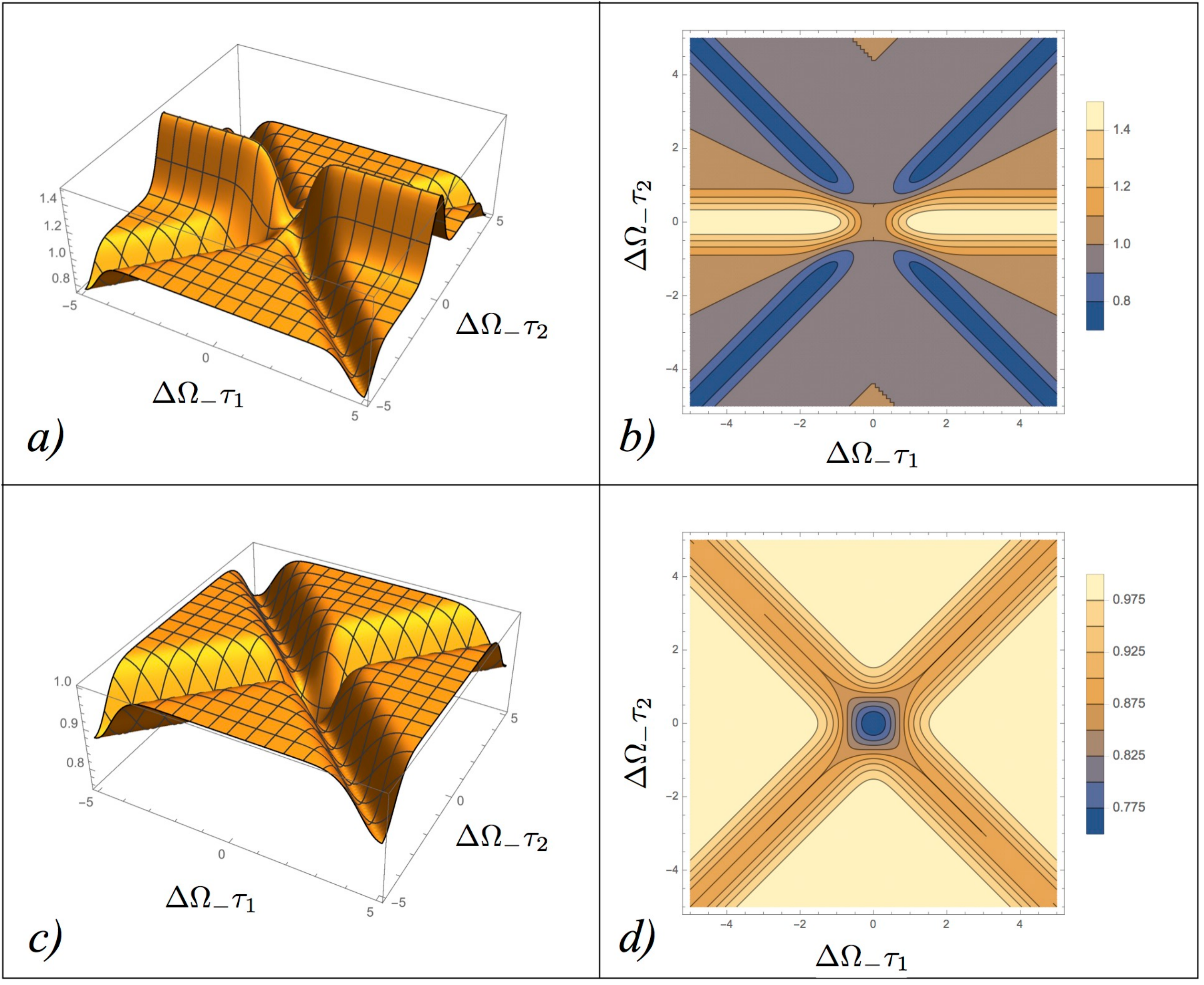} 
	\centering
	\caption{Panel a) and b): 3D  and contour  plots of 
		the coarse grained coincidence counts $\overline{R}_{\mbox{\tiny{BP}}}(\tau_1,\tau_2)$ of
		equation~(\ref{ED1CG}) for the BP pure state $|\Psi^{(a)}_{S} \rangle$, in the generalized HOM scheme. 
		Panel c) and d): 3D  and contour  plots of the coarse grained rates  $\overline{R}_{\mbox{\tiny{CP}}}(\tau_1,\tau_2)$ of equation~(\ref{EP1CG}) associated
		with the CP signal.
		In all the plots the delays times have been rescaled by the inverse of the width $\Delta \Omega_-$  of the bi-photon spectrum, and the coincidence counts have been rescaled by  their plateau values,
		i.e. respectively $\overline{R}_{\mbox{\tiny{BP}}}(0,\infty)=1/2$ and $\overline{R}_{\mbox{\tiny{CP}}}(0,\infty)=A^2$. Besides, setting $\Delta \Omega_+=\Delta \Omega_-/5$ means that $\Delta \omega\simeq \Delta \Omega_-/2$ is applied in c) and d). }
	\label{SS-Fig}
\end{figure*}

The results presented so far shows that our proposed modification of the HOM interferometer allows one to connect two independent spatial coordinates (namely $\Delta \ell_1^{(0)}$ and $\Delta \ell_2^{(0)}$) with the unique zero of the coincidence counts at the output of the setup. Yet we notice that at variance with equation~(\ref{ED1}) that defines the coincidence counts for the standard HOM setting,  the minimum of $R_{\rm BP}(\tau_1,\tau_2)\Big\rvert_{\theta=\pi/2}$ is accompanied by a series of fast oscillations of wavelength $\lambda_0=2\pi c/\omega_0$, similar to those we observed in equation~(\ref{COH}) for the classical inputs in the conventional HOM configuration. For an optical source with $\lambda_0\simeq 300 \; {\rm nm}$ which will produces the interference fringes that, while being observable in the
controlled scenarios~\cite{MZ04,MZ1,MZ2,MZ02,MZ03}, in many practical applications, are typically too small to ensure that the sensitivity of the scheme will not  be compromised. Indeed as we have done in the subsection where reviewing the HOM dip with CP state, under these conditions the measured value of the coincidence counts will be no longer provided by $R_{\mbox{\tiny{BP}}}(\tau_1,\tau_2)$ of equation~(\ref{ED1222}) but by its coarse grained form.  Accordingly, irrespectively from the selected value of $\theta$, we get 
\begin{eqnarray} \label{ED1CG} 
 \overline{R}_{\mbox{\tiny{BP}}}(\tau_1,\tau_2)  
\simeq \frac{1}{8} \Big[  4 +  2 e^{-2\; \Delta^2 \Omega_- \tau_2^2} -e^{-2\; \Delta^2 \Omega_- (\tau_1+\tau_2)^2}-e^{-2\; \Delta^2 \Omega_- (\tau_1-\tau_2)^2}\Big]\;,
\end{eqnarray}
for which not only the origin of the $(\tau_1,\tau_2)$ plane is not a zero, but also it is not even a local minimum, see Figure~{\ref{SS-Fig}}~a) and b).  Even worse than that under the same smoothing, the coincidence counts~(\ref{CP1asdfad}) of the CP input, while being deteriorating,   acquires $(\tau_1,\tau_2)=(0,0)$  as a global minimum, i.e. 
\begin{eqnarray} \label{EP1CG} 
\overline{R}_{\mbox{\tiny{CP}}}(\tau_1,\tau_2) 
\simeq 
A^2 \Big[1 -  \tfrac{ e^{-4\; \Delta^2 \omega(\tau_1+\tau_2)^2} + e^{-4\; \Delta^2 \omega (\tau_1-\tau_2)^2}}{8}\Big]\;, 
\end{eqnarray}
see Figure~{\ref{SS-Fig}} c) and d) (the parameter $A$ being the same as in equation (\ref{AAA}). Does this means that in passing from the original HOM scheme to our modified interferometer no advantage can be expected in using the BP state for sensing in realistic scenarios? As we shall see in the next paragraphs this is not really so: indeed due to the special functional dependence of $\overline{R}_{\mbox{\tiny{BP}}}(\tau_1,\tau_2)$,  even in the presence of coarse graining,  the BP input still provides a better visibility  for two-parameter sensing than the one could achieve using a CP source.  
Before entering in the technical details of the analysis, it is worth stressing that since both~(\ref{ED1CG}) and (\ref{EP1CG})  do not bare any dependence upon the specific value we choose for the  parameter $\theta$, for the discussion that follows the presence of the achromatic phase shifter is completely irrelevant. Accordingly in studying the sensitivity of the scheme in presence of parameter fluctuations we can drop such element, simplifying the design of the interferometer.

\begin{figure}[!t]
	\centering 
	\includegraphics[width=0.5\columnwidth]{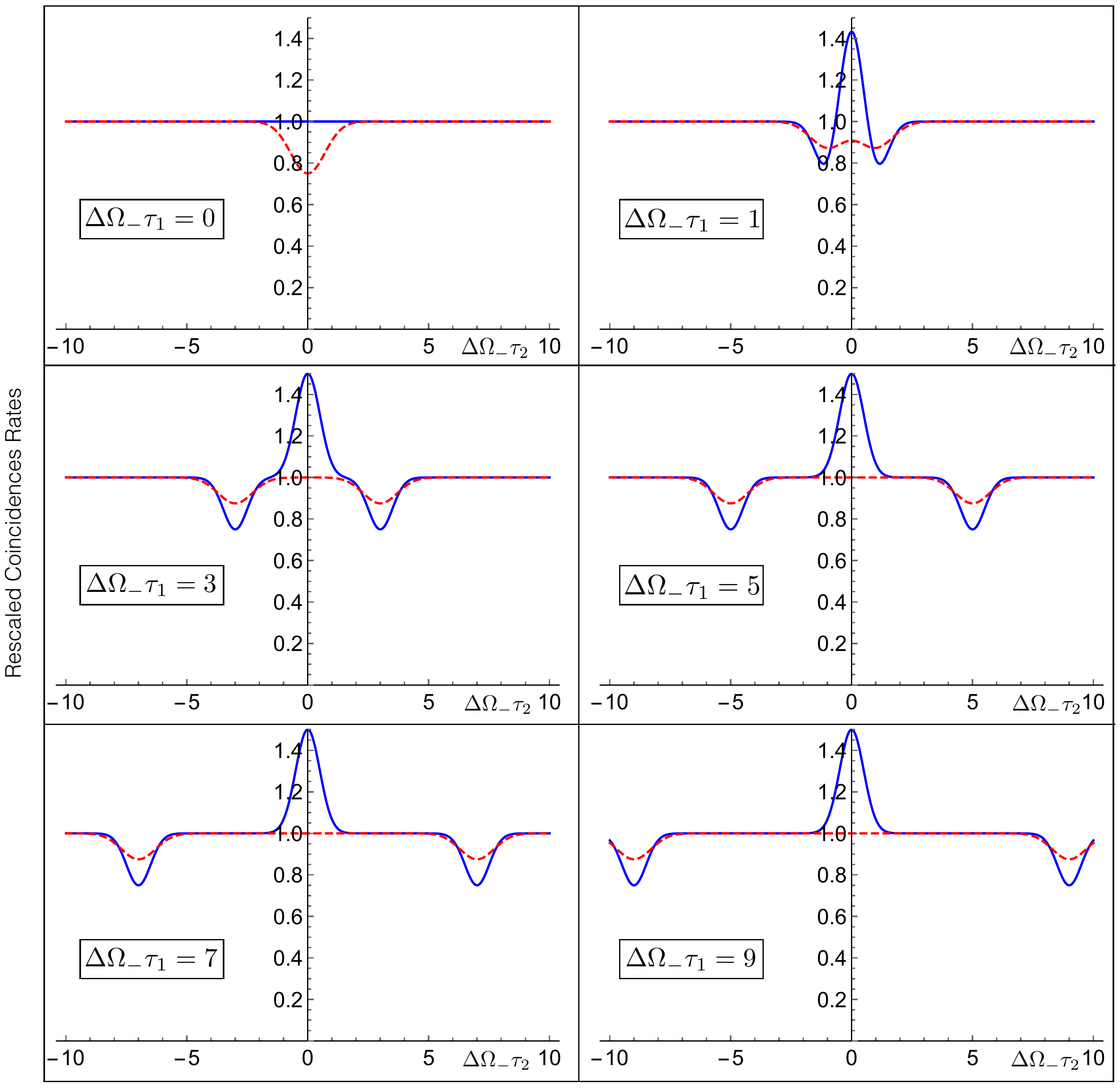} 
	\caption{Functional dependence of the rescaled value of $\overline{R}_{\rm BP}(\tau_1,\tau_2)$ of equation~(\ref{ED1CG}) (blue continuous line) and $\overline{R}_{CP}(\tau_1,\tau_2)$ of equation~(\ref{EP1CG}) 
		(red dashed line) upon $\tau_2$ for assigned values of $\tau_1$. Both of them get the symmetric dips when $\tau_2=\pm\tau_1$, moreover, $\overline{R}_{\rm BP}(\tau_1,\tau_2)$ presents the maximum at $\tau_2=0$. As input state for the CP configuration we assume a
		Gaussian envelop of width $\Delta \omega= \Delta \Omega_-$ corresponding to have  
		$\Delta \Omega_{+}=0$ (i.e. $\Delta \Omega_-\gg \Delta \Omega_+$) in equation~{(\ref{DEFDOMEGA})}.
		As in Figure~\ref{SS-Fig} 
		all coincidences have been rescaled by their corresponding  plateau values, i.e. respectively $\overline{R}_{\mbox{\tiny{BP}}}(\tau_1,\infty)=1/2$ and $\overline{R}_{\mbox{\tiny{CP}}}(\tau_1,\infty)=A^2$.   }
	\label{COM_BP_CP} 
\end{figure}

In Figure~{\ref{COM_BP_CP}  we report the functional dependence of $\overline{R}_{\mbox{\tiny{BP}}}(\tau_1,\tau_2)$ and $\overline{R}_{\mbox{\tiny{CP}}}(\tau_1,\tau_2)$ upon $\tau_2$ for assigned value of $\tau_1$, by rescaling both functions with respect to their $\tau_2=\pm\infty$ asymptotic values (i.e. $\overline{R}_{\mbox{\tiny{BP}}}(\tau_1,\pm\infty)=1/2$ and $\overline{R}_{\mbox{\tiny{CP}}}(\tau_1,\pm\infty)=A^2$). Considering hence the case where  $\tau_1$ is sufficiently large (i.e.  $\Delta\Omega_-\tau_1 > 1$ for the BP case and  $\Delta \omega \tau_1 > 1$ for the CP case) -- if this condition is not met by the parameter $\Delta \ell_1^{(0)}$ simply allow the experimentalist to play with the value of the control term $x_1$ till the condition is enforced. In this regime
$\overline{R}_{\mbox{\tiny{BP}}}(\tau_1,\tau_2)$ presents a maximum at $\tau_2=0$ and two symmetric minima at $\tau_2=\pm \tau_1$, while $\overline{R}_{\mbox{\tiny{CP}}}(\tau_1,\tau_2)$ is everywhere flat with only two symmetric minima at $\tau_2=\pm \tau_1$. Both these features can clearly be used to recover the values of two unknown parameters $\Delta \ell_1$ and $\Delta \ell_2$ by keeping $x_1=0$ constant and swiping with respect to $x_2$. For instance in the case of the CP state, this accounts in recording the values $x_{\min}^{\text{(left)}}$ and $x_{\min}^{\text{(right)}}$ for which the function 
\begin{eqnarray}
f_{\mbox{\tiny{CP}}}(x_2):=\overline{R}_{\mbox{\tiny{CP}}}( \frac{\Delta \ell_1^{(0)}}{2c}, \frac{\Delta \ell_2^{(0)} + x_2}{2c})\;,
\end{eqnarray} 
admits the two minima and then using trivial inversion formulas {$\Delta \ell_1^{(0)}=x_{\min}^{\text{(right)}}-x_{\min}^{\text{(left)}}$} and {$\Delta \ell_2^{(0)}= x_{\min}^{\text{(right)}}+x_{\min}^{\text{(left)}}$}. For the BP state instead one  could employ a different strategy, say recording the position  $x_{\text{max}}$ of the maximum of 
\begin{eqnarray} \label{FBP} 
f_{\mbox{\tiny{BP}}}(x_2):=\overline{R}_{\mbox{\tiny{BP}}}(\frac{\Delta \ell_1^{(0)}}{2c}, \frac{\Delta \ell_2^{(0)} + x_2}{2c})\;,
\end{eqnarray} 
and one of the two minima (say $x_{\min}^{\text{(right)}}$), the inversion formulas being now {$\Delta \ell_1^{(0)}=2(x_{\min}^{\text{(right)}}-x_{\max})$} and {$\Delta \ell_2^{(0)}= 2x_{\max}$}. As in the case of the original HOM setting while both CP and BP provide similar accuracies  that scale as width of the dips, i.e. $c/\Delta \Omega_-$, the special features of $\overline{R}_{\mbox{\tiny{BP}}}$  ensures  a net improvement in the signal visibility due to the fact that both the maximum and the minima of $f_{\mbox{\tiny{BP}}}(x_2)$ are more pronounced with respect to the background signal, than the minima of $f_{\mbox{\tiny{CP}}}(x_2)$: specifically for the BP state  we have $V_{\max}\simeq 50\%$ and $V_{\min}\simeq 25\%$, while for the CP state we have $V_{\min}\simeq 12.5\%$ -- in all the cases $V$ is defined as the percentage of the ratio between the distance of the max (resp. min) from the background (plateau), and the height of the latter. 
\begin{figure*}
		\centering 
		\includegraphics[width=0.5\columnwidth]{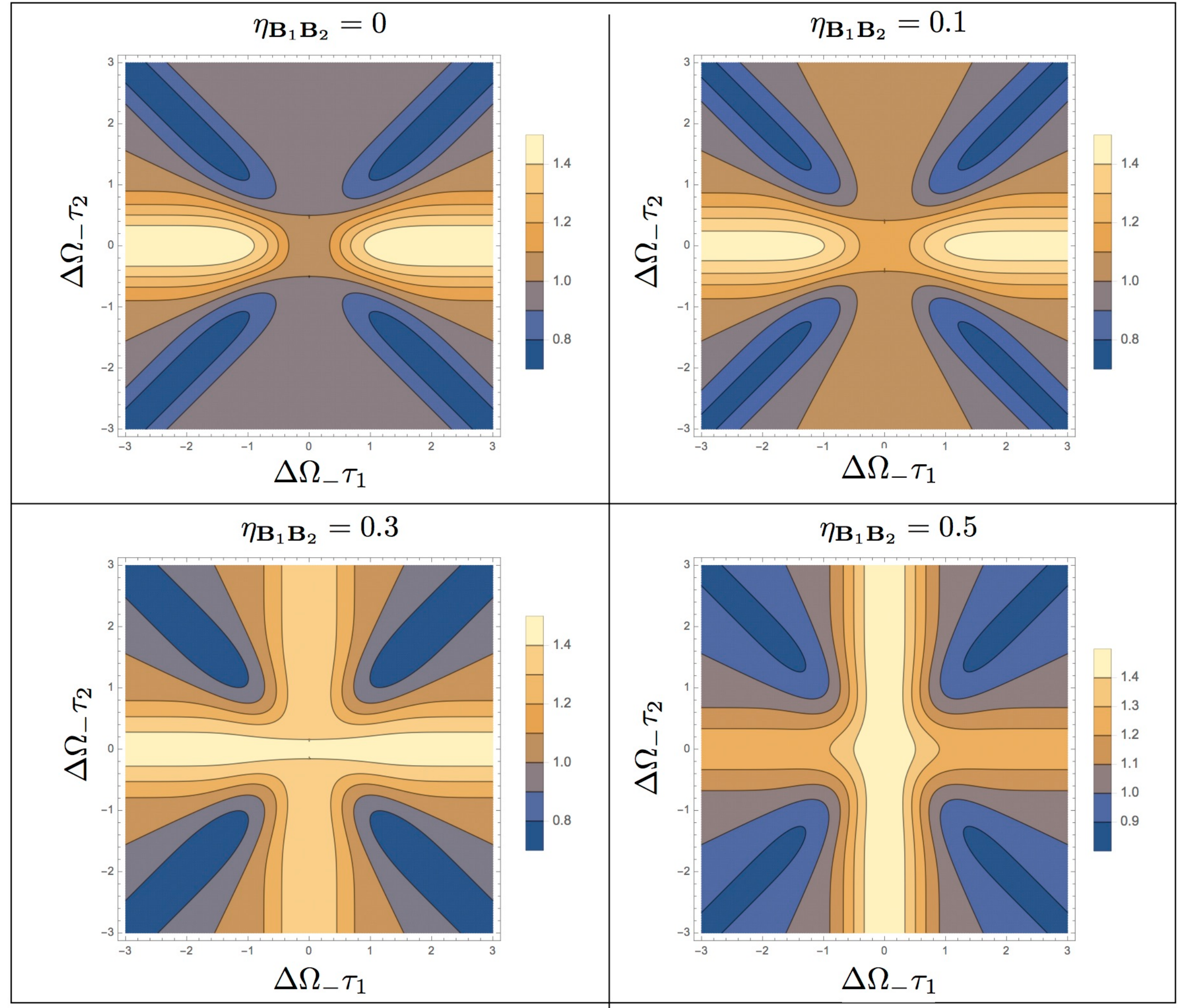} 
		\caption{Contour plots of the coarse grained coincidence counts (\ref{MBP_LOSS}) associated with the BP state for the modified HOM interferometer for different values of the asymmetry loss parameter 
			$\eta_{\mathbf{B}_1\mathbf{B}_2}$ of equation~(\ref{ASB}).
			As usual for the BP configuration we assume the Gaussian frequency amplitude with  $\Delta \Omega_+=\Delta \Omega_-/5$, while the coincidence counts have been rescaled by their
			plateau value $4A_{\rm BP}^{\rm loss}$. 
			\label{MBP_NOISE} }
\end{figure*}
\subsection*{Case 3: losses}
Up to now we considered the case where photons propagate into our modified HOM  setup in the absence of losses.  Nonetheless, as we shall see, the scheme is relatively robust even in the presence of losses, as long as the bi-photon state retains its symmetric character. The analysis which follows enlighten this fact by making use of the formalism in the subsection ``Losses in the HOM interferometer'' of Supplementary Material online we develop  to describe the photon losses in the standard HOM setting. Let us hence  indicate with $\xi_j(\omega)$ and $\chi_j(\omega)$ the absorption amplitudes associated with the path 
$\mathbf{A}_j$ and $\mathbf{B}_j$ of Figure~\ref{Fig1} b), respectively. Similarly to what we have done in the subsection ``Losses in the HOM interference'' of Supplementary Material online for the conventional HOM setting, we shall focus on the white-noise regime, imposing the condition 
\begin{eqnarray} 
\xi_j(\omega) \simeq \xi_j(\omega_0)\;, \qquad \chi_j(\omega) \simeq \chi_j(\omega_0) \label{WITHENOISE1} \;.  
\end{eqnarray}  
As explicitly shown in the section ``Losses for the modified HOM interferometer'' of Supplementary Material online, under this assumption equation~(\ref{ED1CG}) gets replaced by 
\begin{eqnarray}\label{MBP_LOSS}
	\overline{R}_{BP}(\tau_1,\tau_2)&\simeq &A_{\rm BP}^{\rm loss}  
	\Big[  4 +  2 e^{-2\; \Delta^2 \Omega_- \tau_2^2} -e^{-2\; \Delta^2 \Omega_- (\tau_1+\tau_2)^2}-e^{-2\; \Delta^2 \Omega_- (\tau_1-\tau_2)^2}
	\nonumber \\
	&& + \eta_{\mathbf{B}_1\mathbf{B}_2} \Big(   - 2 e^{-2\; \Delta^2 \Omega_- \tau_2^2}+4 e^{-2\; \Delta^2 \Omega_- \tau_1^2} +e^{-2\; \Delta^2 \Omega_- (\tau_1+\tau_2)^2}+e^{-2\; \Delta^2 \Omega_- (\tau_1-\tau_2)^2}
	\Big) \Big]  \;,
\end{eqnarray} 
where $A_{\rm BP}^{\rm loss}$ is a multiplicative factor representing the reduction of intensity of the signal, while 
$\eta_{\mathbf{B}_1\mathbf{B}_2}$ is a parameter that gauges the relative loss mismatch between  the losses associated with the paths $\mathbf{B}_1$ and $\mathbf{B}_2$, i.e.
\begin{eqnarray} 
A_{\rm BP}^{\rm loss} &: =& \tfrac{|\xi_1(\omega_0)\xi_2(\omega_0)|^2(|\chi_1(\omega_0)|^2+ |\chi_2(\omega_0)|^2)^2}{32} \;, \label{ALOSS}  \\
\eta_{\mathbf{B}_1\mathbf{B}_2} &: =&\left( \tfrac{|\chi_1(\omega_0)|^2-|\chi_2(\omega_0)|^2}{|\chi_1(\omega_0)|^2+ |\chi_2(\omega_0)|^2} \right)^2  \;. \label{ASB} 
\end{eqnarray} 
Equation~(\ref{MBP_LOSS}) shows that, as in the original HOM setting,  the unbalance between the losses of input paths $\mathbf{A}_1$ and $\mathbf{A}_2$ does not deteriorate the signal $\overline{R}_{BP}(\tau_1,\tau_2)$. On the contrary as $\eta_{\mathbf{B}_1\mathbf{B}_2}$ approaches $1$, the function loose its dependence upon the parameter $\tau_2$ as clearly shown in Figure~\ref{MBP_NOISE}. Also, at fixed $\tau_1$, the maximum and the minima  of the function~(\ref{MBP_LOSS}) now have a visibility that is reduced by a factor $(1- \eta_{\mathbf{B}_1\mathbf{B}_2})$ with respect to the noiseless case, see also Figure~\ref{Loss-bom}.

We compare these results with the coincidence counts $\overline{R}_{CP}(\tau_1,\tau_2)$ for the  CP input states,  under the same noise assumption of equation~(\ref{WITHENOISE1}). In this case equation~(\ref{EP1CG}) gets replaced by  
\begin{eqnarray}
	\overline{R}_{CP}(\tau_1,\tau_2)&\simeq &A_{\rm CP}^{\rm loss}\left\lbrace 1-D+\eta_{\mathbf{A}_1\mathbf{A}_2}(D-\tfrac{1}2 e^{-4\Delta^2\omega\tau_2^2})\right.\nonumber\\
	&&+\eta_{\mathbf{B}_1\mathbf{B}_2}(D+\tfrac{1}2 e^{-4\Delta^2\omega \tau_1^2})\left.+\eta_{\mathbf{A}_1\mathbf{A}_2}\; \eta_{\mathbf{B}_1\mathbf{B}_2}\left[\tfrac{1}2(e^{-4\Delta^2\omega \tau_2^2}-e^{-4\Delta^2\omega \tau_1^2})-D\right] \right\rbrace \;,{\label{MCP_LOSS}}
\end{eqnarray}
with $D:=(e^{-4\Delta^2\omega (\tau_1-\tau_2)^2}+e^{-4\Delta^2\omega(\tau_1+\tau_2)^2})/8$. And $\eta_{\mathbf{B}_1\mathbf{B}_2}$ is defined as equation ~(\ref{ASB}). Here $ A_{\rm CP}^{\rm loss}$ is the multiplicative factor representing the reduction of intensity and $\eta_{\mathbf{A}_1\mathbf{A}_2}$ gauges the relative difference between the corresponding losses along the paths $\mathbf{A}_1$ and $\mathbf{A}_2$, i.e.
\begin{eqnarray}
A_{\rm CP}^{\rm loss}&:=&\left[A \frac{ (|\chi_1(\omega_0)|^2+|\chi_2(\omega_0)|^2)(|\xi_1(\omega_0)|^2+|\xi_2(\omega_0)|^2)}{4}\right]^2\;, \\
\eta_{\mathbf{A}_1\mathbf{A}_2} &: =&\left( \tfrac{|\xi_1(\omega_0)|^2-|\xi_2(\omega_0)|^2}{|\xi_1(\omega_0)|^2+ |\xi_2(\omega_0)|^2} \right)^2 \;,
\end{eqnarray}
where $A$ being the same as in equation (\ref{AAA}) (the details for this calculation can be found in the final part of  the section ``Losses for the modified HOM interferometer'' of Supplementary Material online). We can see that in this case  the coarse grained coincidence counts of the classical signals is effected by not only the unbalance between the losses of input paths $\mathbf{B}_1$ and $\mathbf{B}_2$, but also the one between $\mathbf{A}_1$ and $\mathbf{A}_2$. In particular we notice that in the high unbalance regime for the losses affecting $\mathbf{A}_1$ and $\mathbf{A}_2$ (i.e. $\eta_{\mathbf{A}_1\mathbf{A}_2}=1$), equation~(\ref{MCP_LOSS}) reduces to 
\begin{eqnarray}
\overline{R}_{CP}(\tau_1,\tau_2)\Big{\rvert}_{\eta_{\mathbf{A}_1\mathbf{A}_2}=1}\simeq A_{\rm CP}^{\rm loss}\Big[ 1 -\frac{1}{2}(1-\eta_{\mathbf{B}_1\mathbf{B}_2})e^{-4\Delta^2\omega\tau_2^2}) \Big]\;,  
\end{eqnarray}
which bares no dependence upon $\tau_1$ -- in the same limit on the contrary (\ref{MBP_LOSS}) maintains explicitly functional dependence upon such delay. 
This difference between equation~(\ref{MBP_LOSS}) and equation~(\ref{MCP_LOSS}) shows that the choice for the biphoton state in our scheme allows a more loose noise condition compared with the classical state.

\begin{figure}
	\centering
	\includegraphics[width=0.5\columnwidth]{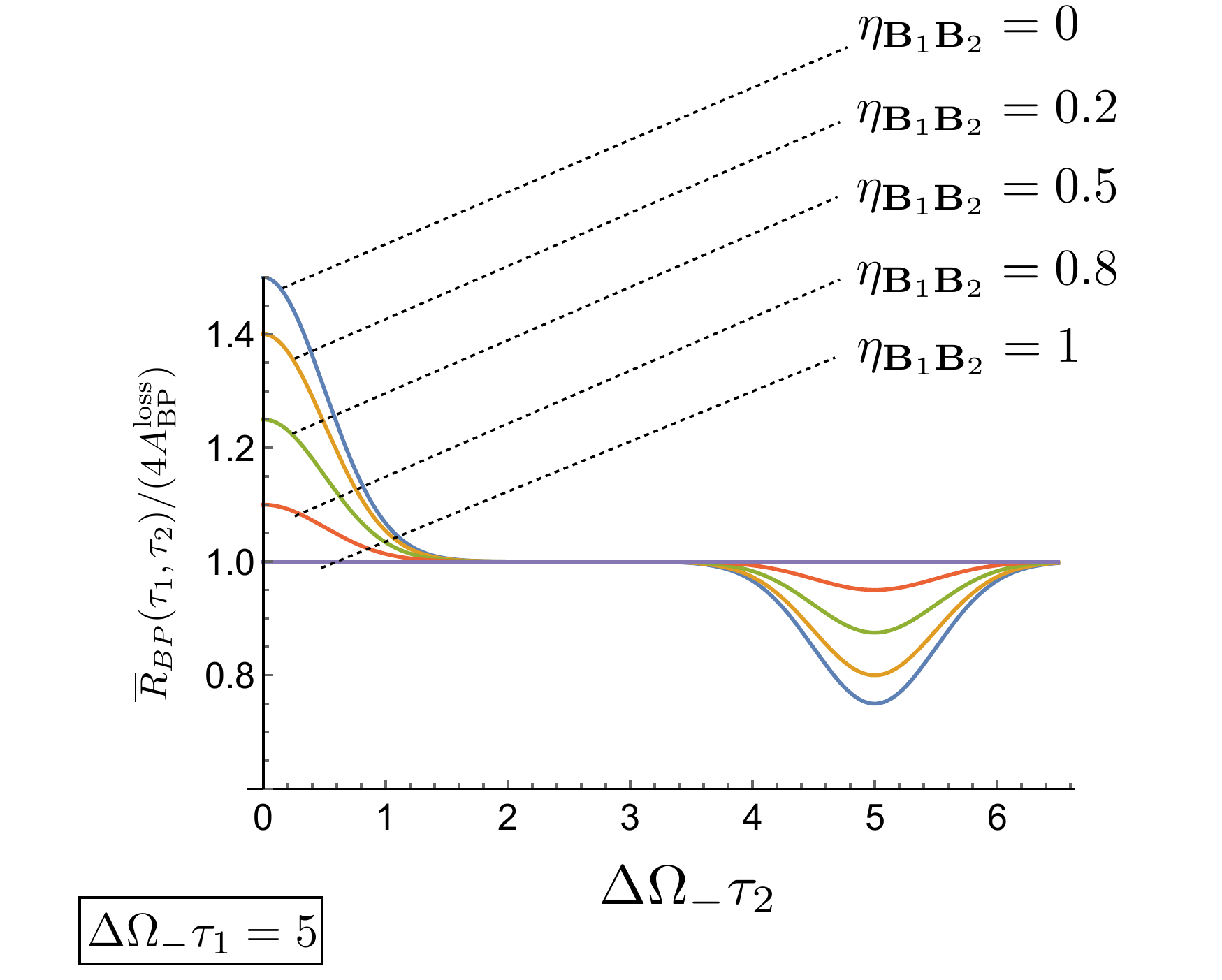}
	\caption{Functional dependence of the coincidence counts $\overline{R}_{\rm BP}(\tau_1,\tau_2)$ of equation~(\ref{MBP_LOSS}) in terms of $\tau_2$ for a fixed value of $\tau_1$ and different values of the
		asymmetry parameter $\eta_{\mathbf{B}_1\mathbf{B}_2}$.}
	\label{Loss-bom}
\end{figure}

\subsection*{Application to QPS} 
In this section we discuss a possible application of our setup in a generalization of the QPS scheme of Ref.~\cite{Bahder:2004}. Such proposal   employed four different baseline setups associated with four independent HOM settings and four dedicated BP sources,  to recover the space-time coordinate of a cooperative  target A provided by a corner cube reflector that allows it bouncing back the impinging optical signals  by  reversing  their propagations directions~\cite{cornereflector}. Basically in the scheme~\cite{Bahder:2004} each of the four coordinates is mapped into a delay of a single HOM and recovered via standard coincidence counts. In our approach of course the same result could be achieved, at least  in principle, with only two BP sources, by employing two independent generalized versions of the HOM interferometer. 

For the sake of simplicity, in what follows we discuss a simplified variant of the positioning task where the location of A is characterized by only two (unknown) independent coordinates. Specifically, as  shown in Figure~\ref{QPSSFIGURE}, we assume A to hover upon the plane where the detectors are located, on a sphere of known radius $r$. This sphere intersects with $y$-axis and $x$-axis at points $R1$, $R2$, $R3$ and $R4$, respectively. The corresponding distance between A and these points are $L1$, $L2$, $L3$ and $L4$. The spatial coordinates of A measured with respect to the cartesian axis indicate can be expressed as $(r\cos\gamma\sin\vartheta, r\cos\gamma\cos\vartheta, r\sin\gamma)$ with angles $\gamma$ and  $\vartheta$. Simple trigonometric consideration allows us to express the distance $L_j$ between A and the points $R_j$ as 
{\small $L1=r\sqrt{2(1+\cos\gamma\cos\vartheta)}$}, {\small $L2=r\sqrt{2(1-\cos\gamma\cos\vartheta)}$}, {\small $L3=r\sqrt{2(1-\cos\gamma\sin\vartheta)}$} and {\small $L4=r\sqrt{2(1+\cos\gamma\sin\vartheta)}$}, respectively. 
The box E at the origin of the coordinate system contains all the optical elements that compose the generalized HOM setting, specifically a bi-photon source, two 50:50 beam splitters, two single-photon detectors. In the ideal case of no fluctuations the box should also contain the achromatic quarter waveplate: in what follows however we shall not consider this possibility assuming instead  the more realistic coarse grained regime~(\ref{CORS}) which does not require the presence of the $\theta=\pi/2$ phase shift. 

\begin{figure}[!t]
	\centering 
	\includegraphics[width=0.5\columnwidth]{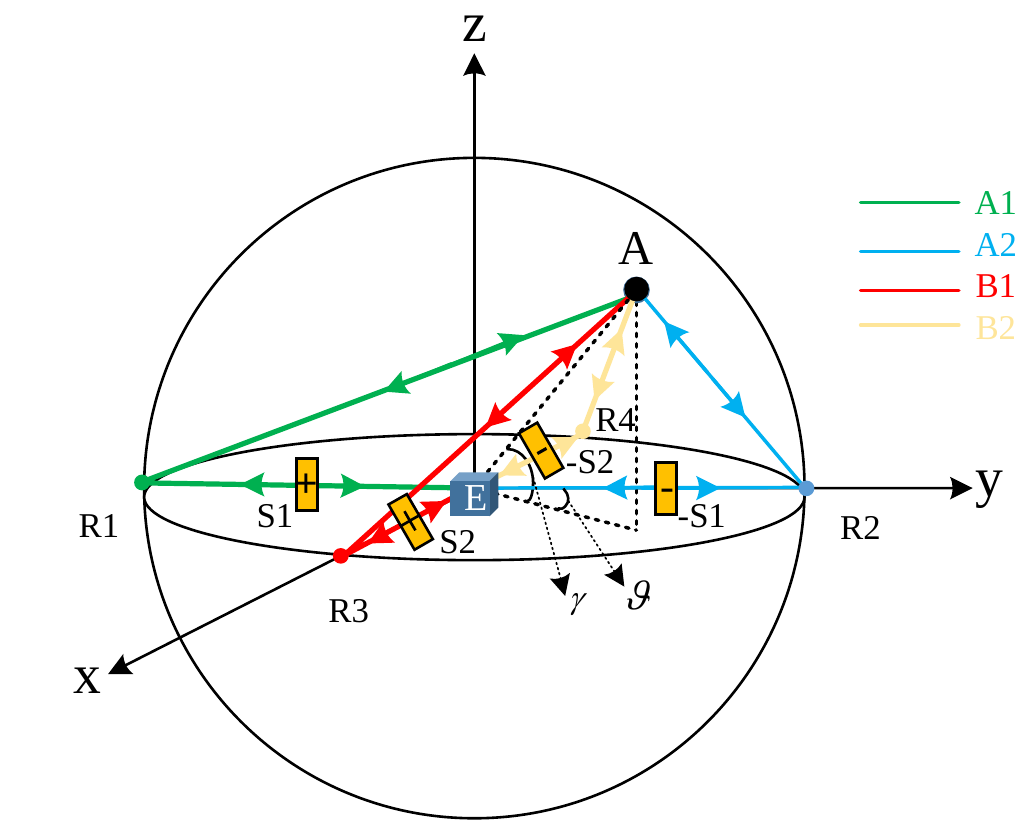}
	\caption{The spatial structure of our QPS scheme for recovering the angular position of a cooperative target A provided by a corner reflector~\cite{cornereflector}. There are two baselines $(R1,R2)$ and $(R3,R4)$ along the y-axis and x-axis, respectively. $S1$, $S2$, $-S1$, and $-S2$ are controllable delays. Optical elements are located in  box E at the origin of the coordinate axis.}
	\label{QPSSFIGURE}
\end{figure}

In order to mimic the modified HOM architecture, the source emits the BP state along the baseline ($R1$, $R2$). After having reached the target  A a first time, the photons are reflected back and interfere through a 50:50 beam splitter placed in E, emerge along the baseline ($R3$, $R4$), reach A a second time, return to E where finally, after being  mixed by a second 50:50 beam splitter, coincidence counts are recorded.   With this choice, as indicated in Figure~\ref{QPSSFIGURE}, the paths $\mathbf{A}_1$, $\mathbf{A}_2$ of the modified HOM scheme can be identified with the trajectories $ER1AR1E$ and $ER2AR2E$ respectively, while  the paths $\mathbf{B}_1$, $\mathbf{B}_2$ with $ER3AR3E$ and $ER4AR4E$. As shown in Figure~\ref{QPSSFIGURE}, we symmetrically put two controllable optical time delays $S1$ and $-S1$ on the baseline $(R1, R2)$, the sign `$-$' expressing  the fact that these two optical delays are adjusted into the opposite directions with the same displacement. Similarly, the time delays $S2$ and $-S2$ are installed on the baseline $(R3, R4)$. Accordingly the zero-delay conditions  is obtained when the identities $L1+S1=L2-S1$ and $L3+S2=L4-S2$  are satisfied, i.e. 
\begin{eqnarray}
&S1=\frac{r}{\sqrt{2}}\left| \sqrt{(1+\cos\gamma\cos\vartheta)}-\sqrt{(1-\cos\gamma\cos\vartheta)}\right|\;,\nonumber \\
&S2=\frac{r}{\sqrt{2}}\left|\sqrt{(1-\cos\gamma\sin\vartheta)}-\sqrt{(1+\cos\gamma\sin\vartheta)} \right| \nonumber\;, 
\end{eqnarray}
which by inversion, allows one to connect the  values of $S_1$ and $S_2$ used to identify the peak and dips of the function~(\ref{FBP}) to the coordinates of the target A.
Clearly in all the transferrings from E to A and from A to E part of the signal will be lost, essentially due to radial dispersion: as discussed about $\overline{R}_{\rm BP}(\tau_1,\tau_2)$ in the above subsection ``Case 3: losses'' this implies that the visibility of the recorded  signal will suffers from the unbalanced between the losses experienced by the photons in their propagation along the baseline  ($R3$, $R4$) (possible asymmetries affecting the baseline ($R1$, $R2$) instead will be uninfluential).

\section*{Conclusion} 
A modified HOM scheme employing an extra 50:50 beam splitter  and an achromatic waveplate to achieve the phase shift in one of the paths has been described in details. By recording the occurrence of a zero coincidence signal at the output of the setting, the distance difference between two paths can be determined. The performance of the scheme has been compared with respect to those one could achieve via intensity-intensity correlation measurement with the classical coherent pulses. When the delays are stable about the wavelength $\lambda_0$ of the mean carrier frequency, the bi-photon scheme exhibits a clear advantage compared to the classical setting (in the latter scenario no zero coincidences can be observed). For unstable delays this is no longer true due to the presence of fast oscillations that tends to wash out the zero coincidence signal: however the coincidences readout associated with a bi-photon source still exhibit a better visibility with respect to one could get from their classical counterpart, which is analogous to what typically observed in the standard HOM scheme. 

We conclude by commenting that, while not explicitly accounted in the present analysis, our setting could be easily adapted to deal with the SPDC  sources which exhibit correlations in the polarization degree of freedom, see e.g. Ref.~\cite{POLA} and references therein.

\section*{Data Availability}
All data analysed during this study are included in this manuscript and its Supplementary Information file.

\section*{Acknowledgements}
The Authors acknowledge fruitful discussions with G. Di Giuseppe, G. C. La Rocca, and D. Vitali. Y. Yang acknowledges support from the Grant National Natural Science Foundation of China 61172138; National Natural Science Foundation of China 61573059; Aviation Science Foundation Project 20160181004; Shanghai Aerospace Science and Technology Innovation Fund SAST2017-030}, and SNS for hosting her.

\section*{Author contributions statement}
Yu Yang achieved all of the calculations and wrote this manuscript,  Luping Xu conceived the application scene of this scheme, and Vittorio Giovannetti analysed the results and improved the efficiency of the scheme presented in this manuscript. All authors reviewed the manuscript.

\section*{Additional Information}
{\bf Supplementary information} accompanies this paper.\\
{\bf Competing Interests}: The authors declare no competing interests.


\begin{thebibliography}{100}
	\bibitem{HOM1} Ou, Z. Y., Hong, C. K. \& Mandel, L. Relation between input and output states for a beam splitter. {\it Opt. Commun.} {\bf 63,} 118-122 (1987).
	
	\bibitem{PhysRevLett.59.2044} Hong, C. K., Ou, Z. Y. \& Mandel, L. Measurement of subpicosecond time intervals between two photons by interference. {\it Phys. Rev. Lett.} {\bf 59,} 2044 (1987).
	
	\bibitem{HOM3} Shih, Y. H. \& Alley, C. O. New type of Einstein-Podolsky-Rosen-Bohm experiment using pairs of light quanta produced by optical parametric down conversion. {\it Phys. Rev. Lett.} {\bf 61,} 2921 (1988).

	\bibitem{MREV} Yuan, Z. S. {\it et al.}  Entangled photons and quantum communication. {\it  Phys. Rep.} {\bf 497,} 1-40 (2010).

	\bibitem{IND1} Steinberg, A. M., Kwiat, P. G. \& Chiao, R. Y.  Dispersion cancellation in a measurement of the single-photon propagation velocity in glass. {\it Phys. Rev. Lett.} {\bf 68,} 2421  (1992).
	
	\bibitem{IND2}   Sergienko, A. V., Shih, Y. H. \& Rubin, M. H. Experimental evaluation of a two-photon wave packet in type-II parametric downconversion. {\it J. Opt. Soc. Am. B} {\bf 12,} 859-862 (1995).
	
	\bibitem{IND5} Kaltenbaek, R., Blauensteiner, B., $\dot Z$ukowski, M., Aspelmeyer, M.  \& Zeilinger, A.  Experimental interference of independent photons. {\it Phys. Rev. Lett.} {\bf 96,} 240502 (2006).
	
	\bibitem{IND3} Bra\'{n}czyk, A. M. Hong-Ou-Mandel interference. Preprint at https://arxiv.org/abs/1711.00080 (2017).
	
	\bibitem{IND4} Jachura, M. \& Chrapkiewicz, R.  Shot-by-shot imaging of Hong-Ou-Mandel interference with an intensified sCMOS camera. {\it Opt. Lett.} {\bf 40,} 1540-1543 (2015).
	
	\bibitem{SPDC2} Grice, W. P. \& Walmsley, I. A.  Spectral information and distinguishability in type-II down-conversion with a broadband pump. {\it Phys. Rev. A} {\bf 56,} 1627 (1997).
	
	\bibitem{SPDC} Shih, Y. Entangled biphoton source - property and preparation. {\it Rep. Prog. Phys.} {\bf 66,} 1009 (2003). 
	
	\bibitem{SPDC1} Rubin, M. H., Klyshko, D. N., Shih, Y. H. \& Sergienko, A. V. Theory of two-photon entanglement in type-II optical parametric down-conversion. {\it Phys. Rev. A} {\bf 50,} 5122 (1994).
	
	\bibitem{SENSOR} Schwarz, L.  \&  van Enk, S. J.  Detecting the drift of quantum sources: not the de finetti theorem. {\it Phys. Rev. Lett.} {\bf 106,} 180501 (2011).
	
	\bibitem{SENSOR1} Lyons A. {\it et al.} Attosecond-resolution Hong-Ou-Mandel interferometry. {\it Sci. Adv.} {\bf 4,} 5 (2018).
	
	\bibitem{CLOCK1} Giovannetti, V., Lloyd, S.  \& Maccone, L.  Quantum-enhanced positioning and clock synchronization. {\it Nature} {\bf 412,} 417-419 (2001).
	
	\bibitem{CLOCK2} Giovannetti, V.,  Lloyd, S., Maccone, L. \& Wong, F. N. C.  Clock synchronization with dispersion cancellation. {\it Phys. Rev. Lett.} {\bf 87,} 117902 (2001).
	
	\bibitem{Bahder:2004}  Bahder, T. B.  Quantum positioning system. Preprint at https://arxiv.org/abs/quant-ph/0406126 (2004).
	
	\bibitem{MZ0} Ou, Z. Y., Zou, X. Y., Wang, L. J.  \& Mandel, L. Experiment on nonclassical fourth-order interference {\it Phys. Rev. A} {\bf 42,} 2957 (1990).
	
	\bibitem{MZ03}  Rarity, J. G. {\it et al.} Two-photon interference in a Mach-Zehnder interferometer. {\it Phys. Rev. Lett.}  {\bf 65,} 1348 (1990).

	\bibitem{MZ04} Larchuk, T. S. {\it et al.}  Interfering entangled photons of different colors. {\it Phys. Rev. Lett.}  {\bf 70,} 1603 (1993).
		
	\bibitem{MZ01}  Lee, H., Kok, P. \& Dowling, J. P. A quantum Rosetta stone for interferometry. {\it J. Mod. Opt.} {\bf 49,} 2325-2338 (2002).
	
	\bibitem{MZ02} Edamatsu, K., Shimizu, R. \& Itoh, T. Measurement of the photonic de Broglie wavelength of entangled photon pairs generated by spontaneous parametric down-conversion. {\it Phys. Rev. Lett.} {\bf 89,} 213601 (2002).

	\bibitem{MZ1} Kim, T., Kim, H., Ko, J. \&  Park, G., Two-photon interference experiment in a Mach-Zehnder interferometer. {\it J. Opt. Soc. Korea} {\bf 7,} 113-118 (2003).
	
	\bibitem{MZ2} Kim, H., Lee, S. M. \&  Moon, H. S., Two-photon interference of temporally separated photons {\it Sci. Rep.} {\bf 6,} 34805 (2016).
	
	\bibitem{ACHRO1} Pancharatnam, S.  Achromatic combinations of birefringent plates-Part II. An achromatic quarter-wave plate. {\it Proc. Indian Acad. Sci.} {\bf 41,} 137-144 (1955).
	
	\bibitem{ACHRO2} Vilas, J. L.,  Sanchez-Brea, L. M.  \&  Bernabeu, E. Optimal achromatic wave retarders using two birefringent wave plates. {\it Appl. Optics} {\bf 52,} 1892-1896 (2013).
	
	\bibitem{GIANNI} Abouraddy, A. F., Yarnall, T. M.  \& Giuseppe, G. D. Phase-unlocked Hong-Ou-Mandel interferometry. {\it Phys. Rev. A} {\bf 87,} 062106 (2013). 
	
	\bibitem{ROY} Glauber, R. J.  Photon correlations. {\it Phys. Rev. Lett.}  {\bf 10,} 84 (1963).
	
	\bibitem{SUD} Sudarshan, E. C. G.  Equivalence of semiclassical and quantum mechanical descriptions of statistical light beams. {\it Phys. Rev. Lett.} {\bf 10,} 277 (1963).
	
	\bibitem{MANDELBOOK} Mandel, L.  \&  Wolf, E.  {\it Optical  Coherence and Quantum Optics} (Cambridge University Press, 1995).
	
	\bibitem{TWIN} Giovannetti, V.,  Maccone, L.,  Shapiro, J. H.  \& Wong, F. N. C. Generating entangled two-photon states with coincident frequencies. {\it Phys. Rev. Lett.} {\bf 88,} 183602 (2002).

	\bibitem{METROLOGY} Giovannetti, V., Lloyd, S. \& Maccone, L. Advances in quantum metrology. {\it Nat. Photonics} {\bf 5,} 222-229 (2011).
	
	\bibitem{cornereflector} Newman, W.  I.   {\it Continuum Mechanics in the Earth Sciences} (Cambridge University Press, 2012). 
	
	\bibitem{POLA} Chen, Y. {\it et al.} Polarization entanglement by time-reversed Hong-Ou-Mandel interference. Preprint at https://arxiv.org/abs/1807.00383 (2018).
\end{thebibliography}
\end{document}